\documentclass[conference]{IEEEtran} 
\IEEEoverridecommandlockouts
\usepackage{cite}
\usepackage{amsmath,amssymb,amsfonts}
\usepackage{algorithmic}
\usepackage{graphicx}
\usepackage{textcomp}
\usepackage{xcolor}
\usepackage{tikz}
\usepackage{float}
\usepackage{pgfplots}
\pgfplotsset{compat=1.18}
\usepackage{cuted}
\usepackage{pgfplots}

\usepackage{subcaption}
\pgfplotsset{compat=1.18}

\def\BibTeX{{\rm B\kern-.05em{\sc i\kern-.025em b}\kern-.08em
    T\kern-.1667em\lower.7ex\hbox{E}\kern-.125emX}}

\usepackage{acronym}
\usepackage[belowskip=-10pt,aboveskip=1.8pt]{caption}
\usepackage{standalone}
\usepackage{subcaption}
\usepackage{tikz}
\usepackage{url,enumitem, cite}
\usepackage{verbatim}
\usepackage[bookmarks,colorlinks]{hyperref}
\usepackage{soul, xcolor}
\usepackage{mathrsfs}
\usepackage[ruled,linesnumbered,vlined]{algorithm2e}
\usepackage{dsfont}
\usepackage{booktabs}
\SetKwInput{KwData}{\textbf{Init}}

\usepackage[all=normal,paragraphs=tight,floats=normal,mathspacing=normal,wordspacing=tight,charwidths=tight,mathdisplays=normal,leading=normal]{savetrees}

\acrodef{lbg}[LBG]{Linde–Buzo–Gray}

\newcommand{\myVec}[1]{{\boldsymbol{#1}}}
\newcommand{\myMat}[1]{{\boldsymbol{#1}}}
\newcommand{\mySet}[1]{\mathcal{#1}}

\let\oldnl\nl
\newcommand{\nonl}{\renewcommand{\nl}{\let\nl\oldnl}}

\newcommand{\figHeight}{0.6\columnwidth}

\newcommand{\Dec}{f_{d}}

\newcommand{\norm}[1]{\left\lVert#1\right\rVert}

\acrodef{dl}[DL]{deep learning}
\acrodef{bs}[BS]{base station}
\acrodef{dnn}[DNN]{deep neural network}
\acrodef{cnn}[CNN]{convolutional neural network}
\acrodef{mlp}[MLP]{multi-layer perceptron}
\acrodef{snr}[SNR]{signal-to-noise ratio}
\acrodef{awgn}[AWGN]{additive white Gaussian noise} 
\acrodef{ml}[ML]{machine learning} 
\acrodef{Adam}[Adam]{stochastic gradient descent} 
\acrodef{mse}[MSE]{mean-squared error}
\acrodef{rmse}[RMSE]{root mean squared error}
\acrodef{rmspe}[RMSPE]{root mean squared periodic error}
\acrodef{mle}[MLE]{maximum likelihood estimation}
\acrodef{snr}[SNR]{signal-to-noise ratio}
\acrodef{admm}[ADMM]{alternating direction method of multipliers}
\acrodef{aoa}[AoA]{Angle of Arrival}
\acrodefplural{aoa}[AoAs]{Angles of Arrival}
\acrodef{em}[EM]{electromagnetic}
\acrodef{rnn}[RNN]{recurrant neural network}
\acrodef{ula}[ULA]{uniform linear array}
\acrodef{em}[EM]{Electromagnetic}
\acrodef{doa}[DoA]{direction of arrival}
\acrodef{music}[MUSIC]{MUltiple SIgnal Classification}
\acrodef{esprit}[ESPRIT]{Estimation of signal parameters via rotational invariance techniques}
\acrodef{evd}[EVD]{eigenvalues decomposition}
\acrodef{sps}[SPS]{spatial smoothing}
\acrodef{iid} [i.i.d] {Independent and identically distributed}
\acrodef{ls}[LS]{Least Square}
\acrodef{relu}[ReLu]{Rectified Linear Unit}
\acrodef{crb} [CRB] {Cramér–Rao Bound}
\acrodef{ccrb} [CCRB] {Conditional Cramér–Rao Bound} 
\acrodef{vqvae}[VQ-VAE]{vector quantized variational autoencoder}
\acrodef{iq}[IQ]{in-phase/quadrature}
\acrodef{esprit}[ESPRIT]{Estimation of Signal Parameters via Rotational Invariance Techniques}
\acrodef{drm}[DR-MUSIC]{Deep Root-MUSIC}
\acrodef{ssn}[SubspaceNet]{Subspace Net}
\acrodef{rssn}[Remote SubspaceNet]{Remote Subspace Net}
\acrodef{uav}[UAV]{unmanned aerial vehicle}
\acrodef{iot}[IoT]{internet-of-things}
\acrodef{ml}[ML]{machine learning}
\acrodef{dl}[DL]{deep learning}
\acrodef{snr}[SNR]{signal-to-noise ratio}
\acrodef{cnn}[CNN]{convolutional neural network}
\acrodef{vit}[ViT]{vision transformer}
\acrodef{dnn}[DNN]{deep neural network}
\acrodef{rf}[RF]{radio frequency}
\acrodef{rff}[RFF]{radio frequency fingerprint}
\acrodef{rffi}[RFFI]{radio frequency fingerprint identification}
\acrodef{gcd}[GCD]{generalized category discovery}
\acrodef{ncd}[NCD]{novel class discovery}
\acrodef{osr}[OSR]{open-set recognition}
\acrodef{vqvae}[VQ-VAE]{vector-quantized variational autoencoder}
\acrodef{vae}[VAE]{Variational Autoencoder}
\acrodef{awgn}[AWGN]{additive white Gaussian noise}
\acrodef{evm}[EVM]{error vector magnitude}
\acrodef{svm}[SVM]{support vector machines}
\acrodef{fc}[FC]{fully connected}
\acrodef{cfo}[CFO]{carrier frequency offset}
\acrodef{iqmm}[IQMM]{IQ mismatch}
\acrodef{fe}[FE]{feature extractor}
\acrodef{lo_leakage}[LO leakage]{local oscillator leakage}
\acrodef{hcf}[HCF]{handcrafted feature}
\acrodef{roc}[ROC]{Receiver Operating Characteristic}
\acrodef{auc}[AUC]{area under curve}
\acrodef{hcf}[HCF]{handcrafted feature}
\acrodef{ce}[CE]{cross-entropy}
\acrodef{knn}[KNN]{K-Nearest Neighbors}
\acrodef{dbscan}[DBSCAN]{Density-Based Spatial Clustering of Applications with Noise}
\acrodef{hdbscan}[HDBSCAN]{Hierarchical Density-Based Spatial Clustering of Applications with Noise}
\acrodef{osrwc}[OSRWC]{open-set-recognition-weighted clustering}
\acrodef{ari}[ARI]{adjusted Rand index}

\title{
{From Identification to Discovery: 
 Holistic RF Fingerprinting  
via Learning-Aided Quantization}
}

\author{
Omer Hazan, Gil Paryanti, and Nir Shlezinger \thanks{
Parts of this work were presented in the 2026 IEEE International Workshop on Signal Processing and AI for Wireless Communications (SPAWC) as the paper~\cite{hazan2026rf}.
O. Hazan and N. Shlezinger are with the ECE School, Ben-Gurion University of the Negev, Be’er-Sheva, Israel (e-mails: hazanom@post.bgu.ac.il;
nirshl@bgu.ac.il).  
 G. Paryanti is with Elbit Systems (email: Gil.Paryanti@elbitsystems.com).  The work was supported by the European Research Council (ERC) under the ERC starting grant nr. 101163973 (FLAIR).}}

\begin{document}

\maketitle


\begin{abstract}
\Ac{rffi} enables device identification and authentication by exploiting transmitter-specific hardware impairments embedded in received RF signals. Existing \ac{rffi} methods are typically designed for a specific operating regime, such as closed-set identification, \ac{osr}, or discovery of new devices, and often treat these tasks as separate algorithmic components. In this paper, we propose a holistic \ac{rffi} framework that jointly supports closed-set identification, \ac{osr}, and \ac{ncd} within a single expandable architecture. The proposed method formulates \ac{rffi} as vector quantization in a learned latent feature space, where registered devices are represented by labeled codewords. This representation enables closed-set identification through codeword assignment, \ac{osr} through uncertainty measures induced by the codebook geometry, and \ac{ncd} by using the latent representations and open-set uncertainty of rejected samples to organize them into new device identities.   We further develop a multi-stage training procedure that combines supervised representation learning, latent-space refinement, codebook initialization, and vector-quantization-aware optimization. Experimental results on LoRa and WiFi \ac{rffi} datasets demonstrate that the proposed framework achieves accurate identification of registered devices, reliable detection of unseen transmitters, and effective discovery and registration of new device classes, highlighting the benefits of coupling the different \ac{rffi} tasks.
\end{abstract}

\acresetall

\section{Introduction}
\label{sec:intro}
Wireless devices are increasingly deployed in dense, heterogeneous, and dynamic environments, ranging from Internet-of-Things networks and industrial monitoring systems to critical communication and sensing infrastructures. This growing reliance on wireless connectivity creates a need for reliable mechanisms for identifying and authenticating transmitters based on their emitted signals~\cite{saadouni2025identification}. Conventional device identification is typically carried out by decoding the transmitted packet and extracting identifiers embedded in  protocol headers. While effective in controlled settings, this approach requires access to the decoded content and assumes compatibility with the employed communication protocol. These requirements may be difficult to satisfy in the presence of encryption, proprietary or heterogeneous protocols, limited receiver resources, or non-cooperative transmissions~\cite{zhang2023radio}.

\Ac{rffi} provides an alternative identification paradigm that operates directly on the received physical-layer signal~\cite{zhang2021radio}. Rather than relying on explicit identifiers encoded in the packet payload or headers, \ac{rffi} exploits device-dependent patterns that are inherently embedded in the analog waveform~\cite{xu2016device,peng2018design}. These patterns arise from unavoidable hardware imperfections in the transmitter chain, such as oscillator offsets, \ac{iq} imbalance, phase noise, nonlinearities, and leakage effects. Since these impairments depend on manufacturing variations and circuit-level properties, they induce RF signatures that are characteristic of individual transmitters~\cite{jian2021radio}. Consequently, \ac{rffi} enables passive and non-cooperative device identification, authentication, and monitoring directly from the observed RF signal~\cite{abbas2024radio}.

The capabilities required from an \ac{rffi} system depend on the relation between the devices observed during deployment and those available during system design. This gives rise to three complementary tasks. The first is \emph{closed-set identification}, where every received signal is assumed to originate from one of the registered devices observed during training. In this case, \ac{rffi} reduces to a supervised classification problem over a fixed device set. The second task is \emph{\ac{osr}}, which relaxes this assumption by requiring the system to recognize when an incoming signal is inconsistent with all registered devices and should therefore be rejected as originating from an unknown transmitter. The third task is \emph{\ac{ncd}}, where signals attributed to unknown transmitters are further organized into previously unseen device identities. When these capabilities are considered jointly, the resulting setting is often referred to as \emph{\ac{gcd}} or open-world recognition~\cite{he2025category}: the system must identify registered devices, detect signals from novel devices, and incorporate the discovered devices into its identification model.

Early \ac{rffi} methods relied primarily on manually designed signal descriptors. These approaches exploit the observation that transmitter-specific hardware impairments manifest themselves in measurable RF attributes, such as \ac{iq} offset~\cite{brik2008wireless}, phase-related features~\cite{hall2003detection}, \ac{iq} imbalance~\cite{huang2017radio}, and error vector magnitude~\cite{candore2009robust}. Such handcrafted features are typically provided to classical machine learning classifiers, yielding interpretable identification mechanisms that are directly tied to physical transmitter impairments. Related statistic-based methods instead characterize devices using global properties of the received signal, such as sample moments and other distributional statistics~\cite{patel2015non}. While these methods highlight the physical origin of RF fingerprints, their performance depends strongly on the chosen features and on the ability of these features to remain discriminative under changing channels, protocols, and acquisition conditions.

In recent years, \ac{dl}-based \ac{rffi} has become the dominant paradigm. \Acp{dnn}, mostly \acp{cnn}~\cite{sankhe2019oracle,shen2021lora} or recurrent architectures~\cite{hu2025lightweight}, have been applied directly to raw \ac{iq} samples or to time-frequency representations, enabling automatic extraction of discriminative RF features.  These methods have substantially improved closed-set identification accuracy, especially when sufficient labeled data are available for the devices of interest. Beyond closed-set classification, \ac{dl} representations have also been used for few-shot, incremental, and open-set variants of specific emitter identification and \ac{rffi}~\cite{wang2022few,wu2023specific,yao2023few,liu2023overcoming,guo2024toward,zhang2025open,li2025fscil,xie2025novel}. Nevertheless, many of these approaches are designed around a particular operating regime, such as identifying a fixed set of devices, detecting unknown devices, or adapting to a small number of new classes.

Several recent works have explicitly considered \ac{osr} and discovery-oriented \ac{rffi} using latent representations learned by neural networks. For \ac{osr}, self-supervised or contrastive representation learning has been combined with distance-based anomaly detection in the latent space~\cite{cao2021open,shen2021lora}. Other approaches use probabilistic or prototype-based modeling, including Gaussian prototypes and Mahalanobis-distance-based detection~\cite{liu2023overcoming}, multi-task prototype learning~\cite{ma2025open}, and siamese comparison for open-set category prediction~\cite{cai2025open}. Additional mechanisms include confidence thresholding with feature enhancement~\cite{luo2025radio} and generative adversarial learning for synthesizing outlier-like signals~\cite{guo2024toward,karunaratne2021open}. For \ac{ncd} and related discovery settings, latent features extracted by neural networks are commonly clustered to organize samples from previously unseen devices into new classes~\cite{vaze2022gcd,wen2023parametric,liu2023overcoming}.

To date, existing \ac{rffi} pipelines usually treat closed-set identification, \ac{osr}, and \ac{ncd} as separate algorithmic components. A representation is first learned for classification, a rejection rule is then introduced to detect unknown devices, and clustering or model update procedures are subsequently applied to the rejected samples. This separation leads to two main limitations: 
$(i)$ methods optimized for closed-set identification do not necessarily provide a representation that can be expanded naturally once new device classes are discovered. Incorporating a new class may therefore require retraining, modifying the classifier, or introducing an additional post-processing mechanism; 
$(ii)$ information accumulated in one task is rarely exploited by the next. In particular, \ac{osr} often produces uncertainty or deviation measures that reflect how an incoming signal differs from the registered-device distribution, yet these measures are typically used only for binary rejection and are not leveraged to guide the discovery of new device identities. 

In this paper, we propose a {\em holistic} \ac{rffi} framework that treats closed-set identification, \ac{osr}, and \ac{ncd} as coupled components of a single open-world identification system. The key idea is to replace the conventional view of \ac{rffi} as classification over a fixed set of devices with a representation-based formulation in which devices are described by a learned and expandable set of codewords.   The codeword structure supports closed-set identification of registered devices, provides a mechanism for registering newly discovered devices by expanding the codebook, and induces uncertainty measures for detecting signals that do not conform to the registered-device set. The latter is not treated merely as a binary rejection decision; rather, it is used to guide the subsequent \ac{ncd} stage, thereby coupling detection of unknown devices with their organization into novel device identities.

The proposed framework is inspired by recent developments in \ac{dl}-aided task-based vector quantization~\cite{fishel2025remote,chen2024information}. Specifically, we formulate \ac{rffi} as vector quantization in a learned latent feature space. A neural feature extractor maps the received signal and informative \acp{hcf} into a latent representation, which is then associated with a learned codebook. Registered devices are represented by labeled codewords. \ac{osr} is performed by evaluating how well the latent representation conforms to the geometry and density structure induced by the registered-device codebook. Samples identified as unknown are then processed by the \ac{ncd} module, which uses both their latent representations and their open-set uncertainty information to discover new device classes and augment the codebook accordingly. 

Our main contributions  are summarized as follows:
\begin{itemize}
\item We introduce a holistic formulation of open-world \ac{rffi} that jointly accounts for closed-set identification, \ac{osr}, and \ac{ncd}. Unlike conventional pipelines that treat these tasks as separate stages, the proposed formulation explicitly transfers information from one task to the next, with particular emphasis on using \ac{osr}-derived uncertainty to guide novel class discovery.

\item We develop a concrete \ac{rffi} algorithm based on learned latent representations and codebook-based decision rules. The algorithm combines feature extraction, closed-set identification, open-set scoring, uncertainty-aware buffering of rejected samples, and \ac{ncd} into a single operational pipeline. The codebook is updated as new device classes are discovered, allowing the identification system to expand without redesigning the overall architecture.

\item We propose a detailed multi-stage training procedure for learning the feature extractor and the initial codebook. The training pipeline combines supervised learning, representation refinement, codebook initialization, and vector-quantization-aware optimization, yielding latent features that are both discriminative for registered devices and suitable for subsequent open-set recognition and discovery.

\item We provide an extensive experimental study on LoRa~\cite{shen2021lora} and WiFi~\cite{hanna2022wisig} \ac{rffi} datasets. The experiments systematically evaluate closed-set identification, \ac{osr}, and \ac{ncd}, and demonstrate the benefit of the proposed holistic design. In particular, the results show that the learned vector-quantized representation supports accurate identification of registered devices, reliable detection of unseen transmitters, and effective discovery and incorporation of new device classes. 
\end{itemize}

The rest of this paper is organized as follows. Section~\ref{sec:system_model} presents the system model and reviews  relevant preliminaries. Section~\ref{sec:method} introduces the proposed method. Section~\ref{sec:experiments} presents our experimental study, and Section~\ref{sec:conclusion} provides concluding remarks.


\section{System Model and Preliminaries}
\label{sec:system_model}

\subsection{Signal Model}
\label{subsec:signal_model}
\ac{rffi} aims to identify and authenticate wireless transmitters by exploiting device-specific characteristics that are embedded in their emitted RF signals. These characteristics originate from inevitable hardware imperfections in the analog transmit chain, such as oscillator offsets, IQ imbalance, nonlinearities, and leakage effects. Since such impairments depend on manufacturing variations and circuit-level properties, they induce transmitter-dependent distortions that can serve as RF fingerprints.

To formulate \ac{rffi}, consider a set of wireless transmitters, where the signal emitted by transmitter $i$ is affected by a device-specific transformation. Let $s(t)$ denote the intended baseband signal at time $t$, and let  $F_i(\cdot)$ represent the transformation induced by the hardware chain of transmitter $i$. The received signal associated with transmitter $i$ is modeled as
\begin{equation}
\label{eqn:Model}
    y(t;i) = h(t) * F_i(s(t)) + n(t)
\end{equation}
where $h(t)$ is the impulse response of the wireless propagation channel,  $n(t)$ denotes additive noise, and $*$ is the convolution operator. Accordingly, the mapping $F_i(\cdot)$ captures the transmitter-dependent RF fingerprint, while $h(t)$ and $n(t)$ represent  effects that are not unique to the device and may vary across time, locations, or acquisition conditions.

\subsection{Problem Formulation}
\label{subsec:Problem}
Given a set of received signal snapshots
${y(t;i)}_{t=1}^{T}$, the objective of \ac{rffi} is to infer the identity of the transmitting device. In practical deployments, however, the set of devices encountered during operation is not necessarily identical to the set of devices available during system design. Consequently, a complete \ac{rffi} system must not only identify previously registered devices, but also recognize when a transmission originates from an unseen device and, eventually, incorporate newly encountered devices into its identification model.

To formalize this setting, let $\mySet{I}$ denote the set of registered devices known during system design. \ac{rffi} tasks include:

\paragraph*{Closed-set identification}
The most common \ac{rffi} setting assumes that every received signal originates from one of the registered devices in $\mySet{I}$. In this case, the goal is to estimate the transmitter identity according to the mapping
\begin{equation}
\label{eqn:CSC}
    \hat{i} = f_{\mySet{I}}^{\rm CSC}\left(\{y(t;i)\}_{t=1}^T\right) \in \mySet{I}
\end{equation}
Closed-set identification therefore reduces \ac{rffi} to a multi-class classification problem over the registered device set.

\paragraph*{\Acf{osr}}
In many scenarios, received signals may originate from devices that are not included in $\mySet{I}$. The objective of \ac{osr} is to determine whether a received signal is consistent with the registered-device distribution or should instead be regarded as originating from an unseen device. We represent the output of the \ac{osr} module as
\begin{equation}
\label{eqn:OSR}
    \hat{\rm P}(i \in \mySet{I}) =
    f_{\mySet{I}}^{\rm OSR}\left(\{y(t;i)\}_{t=1}^T\right) \in [0,1].
\end{equation}
A low value of $\hat{\rm P}(i \in \mySet{I})$ indicates that the observed signal is unlikely to have originated from any registered device.

\paragraph*{\Acf{ncd}}
Once samples from unseen devices have been identified, the next challenge is to discover the underlying device identities without access to their labels. Let ${y(t;i_t)}$ denote a collection of measurements satisfying $i_t \notin \mySet{I}$ for all $t$. The objective of \ac{ncd} is to partition these measurements into groups corresponding to previously unseen devices and to register the discovered device classes. Denoting the set of discovered devices by $\widehat{\mySet{I}}{\rm new}$, the registration process updates the known device set according to
\begin{equation}
    {\mySet{I}} \gets \mySet{I} \cup \widehat{\mySet{I}}_{\rm new}.
    \label{eqn:NCDform}
\end{equation}

 \paragraph*{\Acf{gcd}} The combination of \ac{osr} and \ac{ncd} gives rise to the more general \ac{gcd} setting. In \ac{gcd}, the system is not provided with prior knowledge regarding which samples originate from registered devices and which belong to unseen devices. Instead, it must simultaneously identify registered devices, detect samples associated with novel devices, and organize these samples into new device classes. Consequently, \ac{gcd} represents the open-world formulation of \ac{rffi}.

Our goal is to design a {\em holistic \ac{rffi} algorithm} that carries out all of the above tasks. As \ac{gcd} combines \ac{osr} and \ac{ncd}, we focus on a design that jointly carries out the first three tasks.  
For design purposes, we assume access to labeled data corresponding to measurements from the devices in $\mySet{I}$. This dataset is denoted by
\begin{equation}
    \label{eqn:Data}
    \mySet{D} =
    \left\{
    \big\{y^{(j)}(t)\big\}_{t=1}^T, i^{(j)}
    \right\}_{j=1}^{|\mySet{D}|},
    \quad i^{(j)} \in \mySet{I}.
\end{equation}

\begin{figure}[t]
    \centering
    \includegraphics[
        width=\columnwidth,
        height=\columnwidth,
        keepaspectratio
    ]{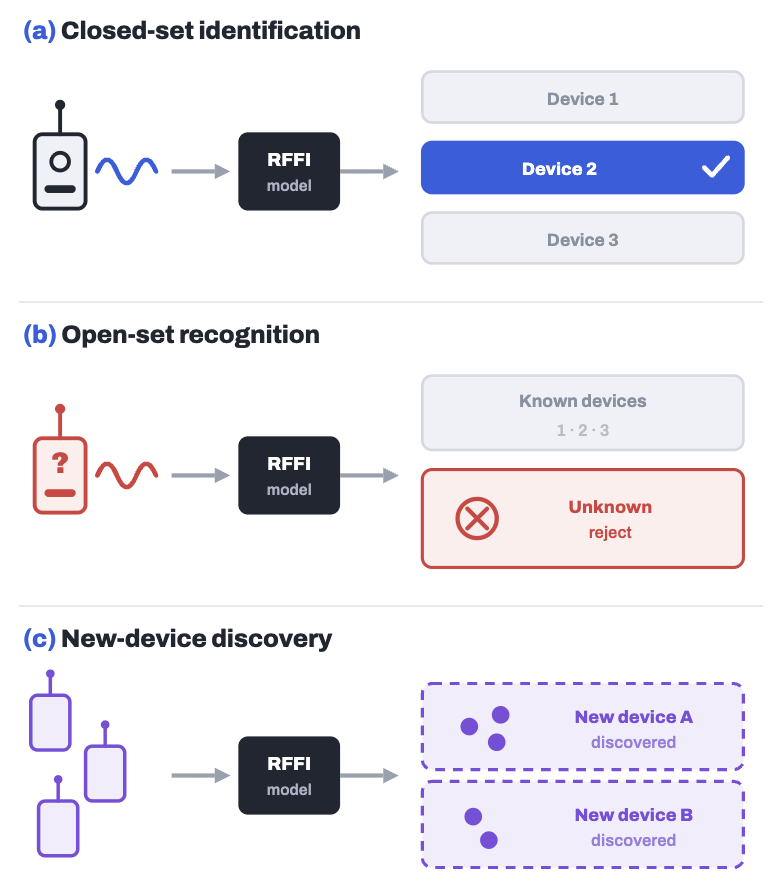}
    \caption{
        Illustration of the three  RFFI tasks:
        closed-set identification, open-set recognition, and novel         class discovery.
    }
    \label{fig:RFFI_tasks}
\end{figure}

\subsection{Preliminaries}
\label{ssec:preliminaries}
Our proposed holistic methodology detailed in Section~\ref{sec:method} is based on tools in vector quantization and its combination with \ac{dl}. We thus briefly review these necessary preliminaries.

\subsubsection{Vector Quantization}
\label{subsec:vector_quantization}
The proposed methodology relies on data-driven vector quantization. A quantizer with input dimension $d$ is defined by a finite codebook $\mySet{Q} \subset \mathbb{R}^d$, such that each input vector $\myVec{x}\in \mathbb{R}^d$ is mapped into a $B$-bit representation $Q(\myVec{x})\in\mySet{Q}$, where $B=\log_2 |\mySet{Q}|$. This mapping is typically performed by selecting the nearest codeword in $\mySet{Q}$~\cite[Ch. 23]{polyanskiy2014lecture}.

When $d>1$, this process is referred to as vector quantization, in contrast to scalar quantization, where $d=1$. Rate-distortion theory shows that, under appropriate assumptions, vector quantization can achieve lower distortion than scalar quantization for a given bit budget~\cite{polyanskiy2014lecture}. A classical method for learning the codebook $\mySet{Q}$ from data is the \ac{lbg} algorithm, in which the codewords are iteratively updated as centroids of clustered data points~\cite{linde1980algorithm}.

\subsubsection{\ac{dl}-Aided Vector Quantization}
\label{subsec:dl_aided_vq}
\noindent
Recent \ac{dl}-based methods have demonstrated the ability to learn multivariate quantized representations from data. A representative example is the \ac{vqvae}~\cite{van2017neural}. In a \ac{vqvae}, an input vector $\myVec{x}$ is first mapped by a \ac{dnn} encoder into a latent representation
    $\myVec{x}^{\rm e} = f_{\rm e}(\myVec{x})$.
The latent representation is then decomposed into $M$ vectors of size $d$, denoted by $\{\myVec{x}_{m}^{\rm e}\}_{m=1}^M$, and each vector is quantized by selecting the closest codeword in $\mySet{Q}$:
\begin{equation}
    \label{eqn:quant2ind}
    \myVec{q}^{m} =
    \mathop{\arg\min}_{\myVec{q}\in\mySet{Q}}
    \|\myVec{x}_{m}^{\rm e} - \myVec{q}\|_2.
\end{equation}
The full quantized representation is therefore
\begin{equation}
    \myVec{q} = Q(\myVec{x}^{\rm e}) =
    [\myVec{q}^1,\ldots,\myVec{q}^M] \in \mySet{Q}^M,
\end{equation}
and is represented using $M \cdot B$ bits. A decoder then maps the quantized representation into the output
   $\hat{\myVec{x}}=\Dec(\myVec{q})$.

To jointly train the encoder, decoder, and codebook, the \ac{vqvae} commonly uses a three-term loss function~\cite{van2017neural}:
\begin{align}
\mathcal{L}^{\rm tot}(\myVec{x}) =
&\mathcal{L}(\myVec{x};\hat{\myVec{x}})
+ \norm{{\rm sg}\left(\myVec{x}^{\rm e}\right)-\myVec{q}}_{2}^{2}
\notag \\
&
+ \beta
\norm{\myVec{x}^{\rm e}-{\rm sg}\left(\myVec{q}\right)}_{2}^{2},
\label{eqn:train_loss}
\end{align}
where ${\rm sg}(\cdot)$ is the stop-gradient operator, $\mathcal{L}(\myVec{x};\hat{\myVec{x}})$ is a task-dependent reconstruction or prediction loss, and $\beta>0$ is a hyperparameter. Alternative loss functions have also been proposed to improve codebook utilization~\cite{fifty2024restructuring} and to support remote inference~\cite{malka2024decentralized}. Nevertheless, the loss in \eqref{eqn:train_loss} remains a common training objective for \ac{dnn}-aided vector quantizers~\cite{huang2023towards}.

\subsubsection{Density-Based Clustering}
\label{subsec:density_clustering}

Density-based clustering identifies clusters as densely populated regions separated by sparse regions. In \ac{dbscan}~\cite{ester1996density}, the $\varepsilon$-neighborhood of a sample $\myVec{z}$ is defined as
\begin{equation}
    \mySet{N}_{\varepsilon}(\myVec{z})
    =
    \left\{
    \myVec{z}'
    :
    \left\|
    \myVec{z}-\myVec{z}'
    \right\|_2
    \leq
    \varepsilon
    \right\}.
\end{equation}
A sample is termed a core point if
$|\mySet{N}_{\varepsilon}(\myVec{z})|\geq m_{\rm min}$, where
$m_{\rm min}$ is the minimum neighborhood size. Clusters are formed by
connecting density-reachable core points and their neighboring samples,
while samples that are not associated with a sufficiently dense region
are labeled as noise.

\ac{hdbscan} extends this principle by constructing a hierarchy of
density-based clusterings rather than relying on a single neighborhood
radius~\cite{campello2013density}. For a selected neighborhood size
$m_{\rm min}$, the core distance of $\myVec{z}$ is defined as the
Euclidean distance to its $m_{\rm min}$th nearest neighbor:
\begin{equation}
    d_{\rm core}(\myVec{z})
    =
    \left\|
    \myVec{z}
    -
    \myVec{z}_{(m_{\rm min})}
    \right\|_2,
    \label{eqn:hdbscan_core_distance}
\end{equation}
where $\myVec{z}_{(m_{\rm min})}$ denotes the $m_{\rm min}$th nearest
neighbor of $\myVec{z}$. For two samples $\myVec{z}_a$ and
$\myVec{z}_b$, let
   $ d_{ab}
    =
    \left\|
    \myVec{z}_a-\myVec{z}_b
    \right\|_2$.
Their mutual-reachability distance is
\begin{equation}
    d_{\rm mr}(\myVec{z}_a,\myVec{z}_b)
    =
    \max\left\{
    d_{\rm core}(\myVec{z}_a),\,
    d_{\rm core}(\myVec{z}_b),\,
    d_{ab}
    \right\}.
    \label{eqn:mutual_reachability}
\end{equation}
Thus, samples in sparse regions are effectively separated by larger
distances. \ac{hdbscan} constructs a hierarchy using these distances and
extracts persistent dense regions as clusters, while retaining
insufficiently supported samples as noise.

\section{Holistic RF Fingerprinting Method}
\label{sec:method}
The proposed method is designed to realize \ac{rffi} as an adaptive open-world identification system, rather than as a collection of separate modules for closed-set classification, \ac{osr}, and \ac{ncd}. 
%
To enable such coupling, we formulate \ac{rffi} through a {\em learned vector-quantized latent representation}. The received signal is mapped into a discriminative feature space, in which registered devices are represented by labeled codewords.  This representation is compact, informative, and expandable:  closed-set identification is carried out by assigning a sample to the device associated with its closest codeword; 
\ac{osr} is performed by evaluating the relation between the sample and the codebook geometry; and newly discovered devices can be incorporated by augmenting the codebook, without redesigning the feature extractor or retraining the entire identification architecture. 

The holistic aspect of the proposed framework follows from the way information flows between the different \ac{rffi} tasks. The codebook learned for registered devices defines the closed-set identification rule and induces uncertainty measures for open-set recognition. 
Thus, the proposed method, whose operation and training procedure are detailed in Subsections~\ref{subsec:rffi_algorithm}-\ref{subsec:rffi_training}, respectively, couples \ac{osr} and \ac{ncd} within a single adaptive \ac{rffi} pipeline, where the output of one task informs the next.

\subsection{RFFI Algorithm}
\label{subsec:rffi_algorithm}

\begin{figure}[t]
    \centering
    \includegraphics[
        width=\columnwidth,
        keepaspectratio
    ]{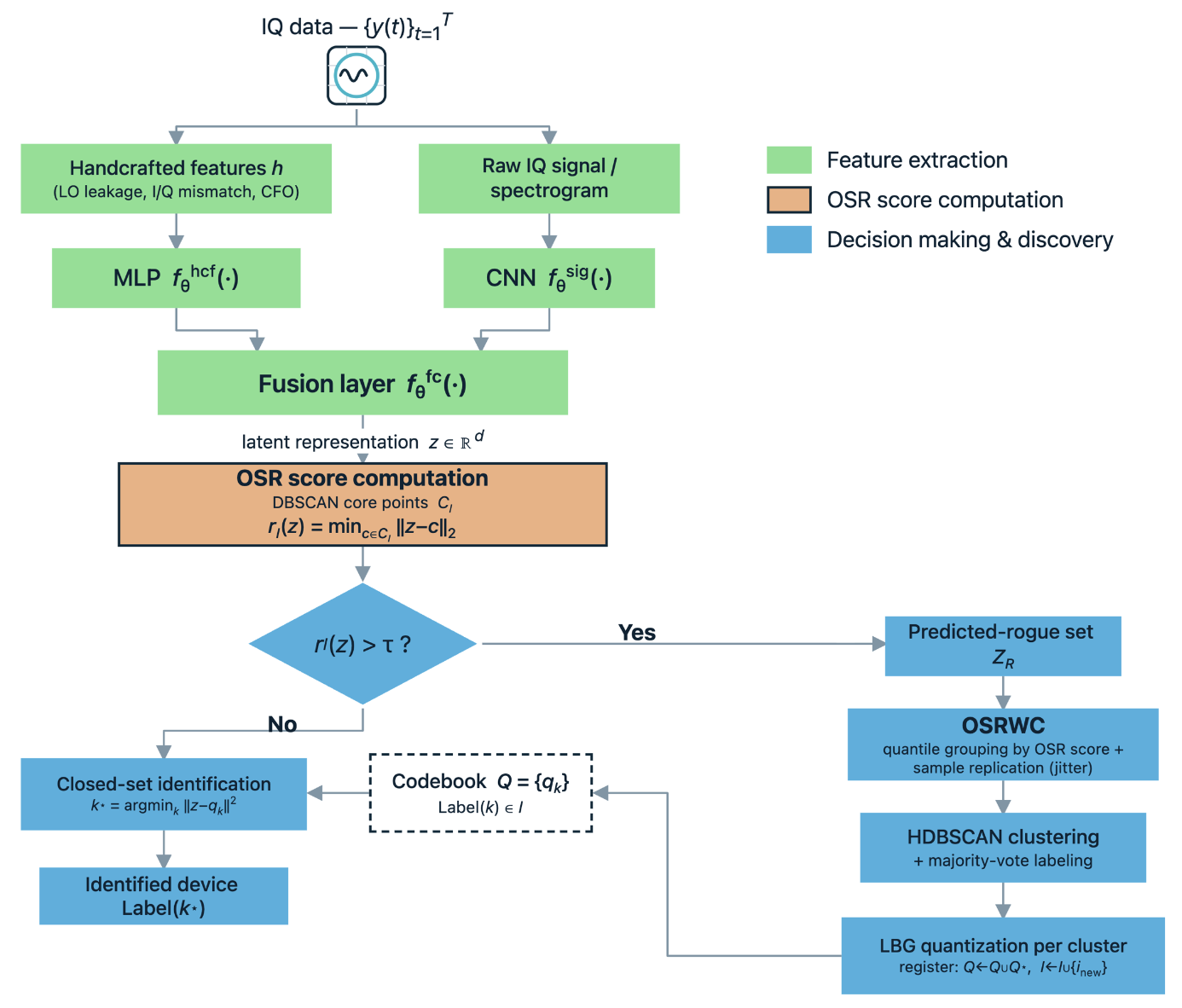}
    \caption{
        Block diagram of the proposed holistic \ac{rffi} algorithm.
    }
    \label{fig:rffi_block_diagram}
\end{figure}

Our proposed \ac{rffi} algorithm operates on a sequence of received IQ
snapshots denoted by $\{y(t)\}_{t=1}^{T}$, and is designed to jointly
perform closed-set identification, \ac{osr}, and \ac{ncd}. The algorithm,
illustrated in Fig.~\ref{fig:rffi_block_diagram}, is composed of three
stages: {\em feature extraction}, {\em OSR score computation}, and
{\em decision making}. The first stage maps the received signal into a
continuous latent representation. The second evaluates the conformity
of this representation with the registered device set, represented by a discrete latent space. Finally, the
resulting latent representation and continuous \ac{osr} score are
jointly used for identification, rogue-device detection, and
new-device discovery.

\subsubsection{Feature Extraction}
\label{subsec:feature_extraction}

The first stage maps the received signal into a continuous latent
representation through two parallel feature-extraction branches. The
first branch extracts \acp{hcf} directly from the received signal. These
features capture device-dependent RF impairments and include
\ac{lo_leakage}, amplitude mismatch, and \ac{cfo}. Denoting the
resulting feature vector by $\myVec{h}$, it is processed by a \ac{mlp} denoted by
$f_{\myVec{\theta}}^{\rm hcf}(\cdot)$.
The second branch processes the received signal representation using a
\ac{cnn}. The input to this branch
can be the raw IQ signal or signal-specific pre-processing (e.g., a spectrogram). We denote
this signal-processing branch by
$f_{\myVec{\theta}}^{\rm sig}(\cdot)$.

The outputs of the two branches are concatenated and processed by an
additional trainable layer into the   representation
\begin{equation}
    \myVec{z}
    =
    f_{\myVec{\theta}}^{\rm fc}
    \left(
    \left[
    f_{\myVec{\theta}}^{\rm sig}
    \left(
    \{y(t;i)\}_{t=1}^{T}
    \right),
    f_{\myVec{\theta}}^{\rm hcf}
    \left(
    \myVec{h}
    \right)
    \right]
    \right)
    \in \mathbb{R}^{d},
    \label{eqn:feature_extractor}
\end{equation}
where $d$ is the latent dimension. Accordingly,
$\myVec{\theta}$ accommodates the trainable parameters of the \ac{cnn},
the \ac{mlp} applied to the \acp{hcf}, and the fusion layer. The
latent vector $\myVec{z}$ serves as a continuous representation of the
device-specific signal characteristics and is provided to the
subsequent \ac{osr} score computation stage.

\subsubsection{\ac{osr} Score Computation}
\label{subsec:rogue_detection}

Rogue-device detection in \ac{rffi} is an instance of the standard
\ac{osr} problem in classification. While closed-set identification
assigns each received signal to one of the registered devices, the
open-set mechanism must determine whether the corresponding IQ data is
consistent with the data observed from the registered set. A signal
that does not conform to this known-device representation is rejected
and regarded as originating from a rogue device. 
To characterize the regions occupied by registered devices in the
continuous latent space, we apply \ac{dbscan}, reviewed in
Subsection~\ref{subsec:density_clustering}, to the representations
produced by the feature extractor. Let
\begin{equation}
    \mySet{Z}_{\mySet{I}}
    =
    \left\{
    \myVec{z}^{(j)}
    =
    f_{\myVec{\theta}}
    \left(
    \{y^{(j)}(t)\}_{t=1}^{T}
    \right)
    \right\}_{j=1}^{|\mySet{D}|}
    \label{eqn:known_latent_set}
\end{equation}
denote the latent representations of the registered-device dataset.
The \ac{dbscan} parameters $\varepsilon$ and $m_{\rm min}$ are selected
through a grid search over registered-device data. The search is
restricted to configurations for which the number of discovered
clusters equals the number of registered devices:
\begin{equation}
    N_{\rm clust}
    \left(
    \operatorname{DBSCAN}
    \left(
    \mySet{Z}_{\mySet{I}};
    \varepsilon,m_{\rm min}
    \right)
    \right)
    =
    |\mySet{I}|.
    \label{eqn:dbscan_cluster_constraint}
\end{equation}
Among the parameter pairs satisfying
\eqref{eqn:dbscan_cluster_constraint}, we select the configuration that
provides the highest coverage of registered-device samples and,
consequently, the lowest rejection rate. 

After fitting \ac{dbscan} using the selected parameters, we retain its
set of core points, denoted by
\begin{equation}
    \mySet{C}_{\mySet{I}}
    =
    \operatorname{Core}
    \left(
    \operatorname{DBSCAN}
    \left(
    \mySet{Z}_{\mySet{I}};
    \varepsilon^{\star},m_{\rm min}^{\star}
    \right)
    \right).
    \label{eqn:dbscan_core_set}
\end{equation}
For an incoming signal with latent representation $\myVec{z}$, we
define the continuous open-set uncertainty score as its Euclidean
distance from the closest registered-device core point:
\begin{equation}
    r_{\mySet{I}}(\myVec{z})
    =
    \min_{\myVec{c}\in\mySet{C}_{\mySet{I}}}
    \left\|
    \myVec{z}-\myVec{c}
    \right\|_2.
    \label{eqn:osr_core_distance}
\end{equation}
A small value of $r_{\mySet{I}}(\myVec{z})$ indicates proximity to a
well-supported registered-device region, whereas a large value
indicates stronger deviation from the registered set.

\begin{figure*}
    \centering
    \includegraphics[
        width=0.8\linewidth,
        trim={0.25cm 0.15cm 0.25cm 0.15cm},
        clip
    ]{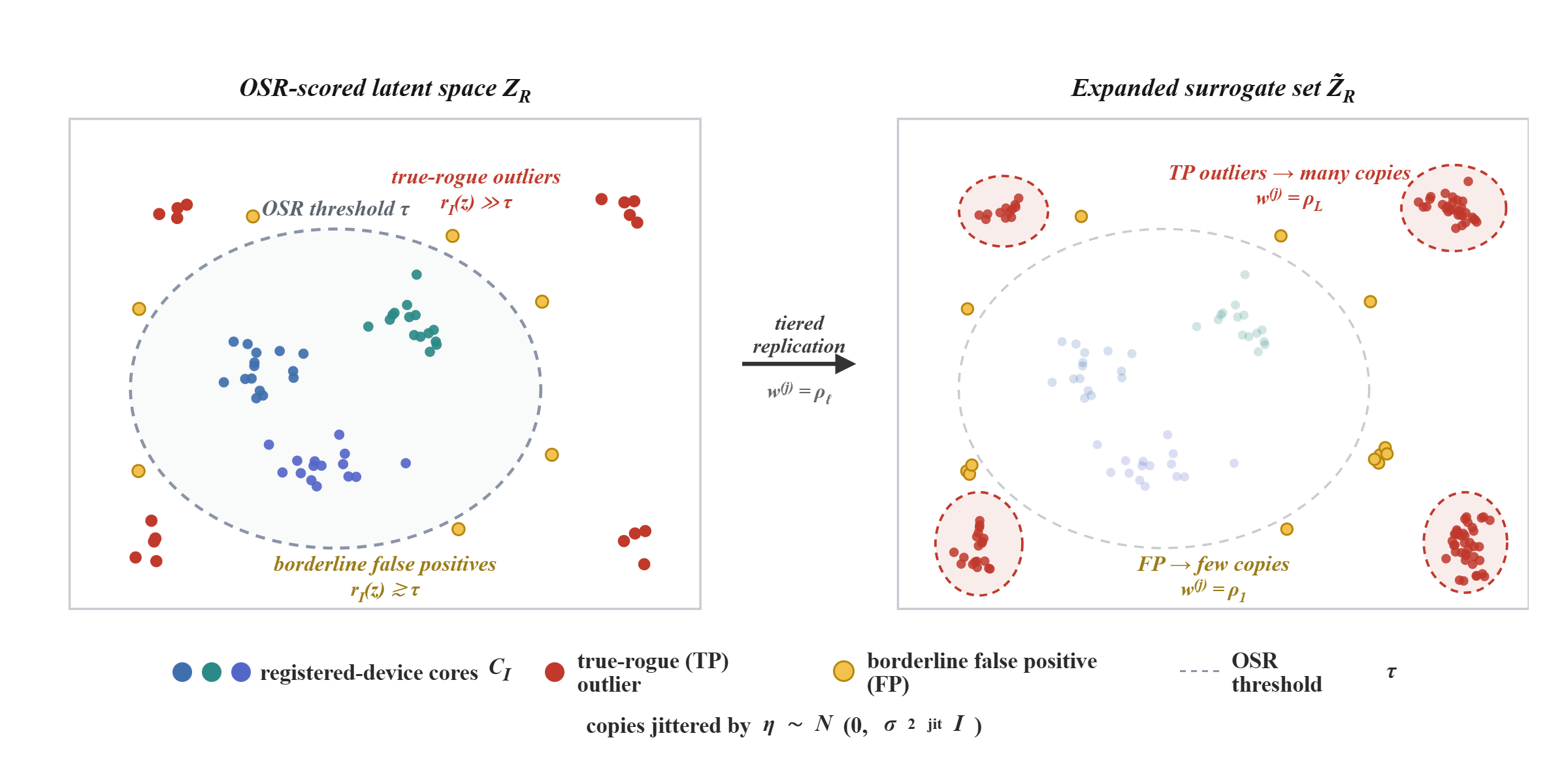}
    \caption{
        Illustration of \acs{osrwc}. High-confidence rogue samples
        receive more jittered copies, increasing their local density
        for \ac{hdbscan}, while borderline false positives receive
        fewer copies.
    }
    \label{fig:osrwc}
\end{figure*}

\subsubsection{Decision Making}
\label{subsec:decision_making}

The first two stages associate each received signal with the pair $\left(\myVec{z},r_{\mySet{I}}(\myVec{z})\right)$, consisting of its continuous latent representation and its deviation from the registered-device regions. The decision stage uses this pair both for the instantaneous identification decision and, when the signal is rejected, for subsequent novel-class discovery. This retention of the continuous \ac{osr} information is central to the holistic nature of the proposed approach.

We first determine whether the signal is consistent with the registered device set by comparing its \ac{osr} score with a threshold $\tau$:
\begin{equation}
\hat{o}(\myVec{z}) =
\mathds{1} \left\{ r_{\mySet{I}}(\myVec{z})>\tau \right\},
\label{eqn:rogue_decision}
\end{equation}
where $\mathds{1}\{\cdot\}$ is the indicator function. Thus, $\hat{o}(\myVec{z})=0$ indicates a registered-device decision and $\hat{o}(\myVec{z})=1$ indicates a rogue-device decision. The threshold $\tau$ is calibrated using registered-device data, such that the open-set detector does not require examples from devices that may be encountered only after deployment.

When $\hat{o}(\myVec{z})=0$, identification is performed using the learned vector-quantized representation. Let $\mySet{Q}=\{\myVec{q}_k\}_{k=1}^{P}$ denote the codebook representing the registered devices, and let $\operatorname{Label}(k)\in\mySet{I}$ denote the device associated with the $k$th codeword. The estimated identity is
\begin{equation}
f_{\mySet{I}}^{\rm CSC}\left( {y(t)}_{t=1}^{T} \right)
= \operatorname{Label}\left(
\mathop{\arg\min}_{k\in{1,\ldots,P}}
\left|\myVec{z}-\myVec{q}_k \right|_2^2 \right).
\label{eqn:codebook_classification}
\end{equation}
Since multiple codewords can be associated with a single device, this representation can accommodate intra-device variability while retaining a simple nearest-codeword identification rule.

When $\hat{o}(\myVec{z})=1$, the latent representation is instead retained for \ac{ncd}, together with its continuous open-set score. Accordingly, over a discovery interval we accumulate
\begin{equation}
\mySet{B} = \left\{ \left(
\myVec{z}^{(j)},r^{(j)} \right) :
r^{(j)} = r_{\mySet{I}} \left( \myVec{z}^{(j)} \right) > \tau \right\}.
\label{eqn:rogue_buffer}
\end{equation}
Note that $\mySet{B}$ may contain both samples from genuinely unseen devices and registered-device samples that were erroneously rejected. A conventional sequential approach would discard the values $r^{(j)}$ after the binary decision in \eqref{eqn:rogue_decision} and cluster all rejected samples equally. Here, we exploit these scores as information transferred from \ac{osr} to \ac{ncd}.

Specifically, we propose {\em \ac{osrwc}}, which uses the magnitude of $r^{(j)}$ as a soft indication of how strongly the $j$th rejected sample departs from the registered-device distribution, as illustrated in Fig.~\ref{fig:osrwc}. We partition the interval $(\tau,\infty)$ according to score quantiles
$\tau=\gamma_0<\gamma_1<\cdots<\gamma_L=\infty$,
and associate the intervals with nondecreasing integer weights $1\leq\rho_1\leq\cdots\leq\rho_L$. The weight assigned to a rejected sample is
\begin{equation}
w^{(j)} = \rho_{\ell},
\qquad
\gamma_{\ell-1} < r^{(j)} \leq \gamma_{\ell}.
\label{eqn:osrwc_weight}
\end{equation}
Thus, samples lying only slightly beyond the rejection boundary have limited influence, whereas samples with stronger evidence of being outside the registered-device regions have larger weights.

We transfer these weights to the density-based discovery stage through local replication. In particular, we construct
\begin{equation}
\widetilde{\mySet{Z}}_{\rm R} =
\bigcup_{(\myVec{z}^{(j)},r^{(j)})\in\mySet{B}} \left\{ \myVec{z}^{(j)}+\myVec{\eta}_{j,m} \right\}_{m=1}^{w^{(j)}}, 
\label{eqn:osrwc_surrogate_set}
\end{equation}
with $\myVec{\eta}_{j,m} \sim \mathcal{N} \left( \myVec{0}, \sigma_{\rm jit}^{2}\myMat{I} \right)$, where $\sigma_{\rm jit}^{2}$ is chosen sufficiently small such that the replicas remain in the immediate neighborhood of the original sample.
The small perturbation avoids coincident points while preserving their location in the latent space.

This construction provides the mechanism through which information obtained in \ac{osr} affects \ac{ncd}. As reviewed in Subsection~\ref{subsec:density_clustering}, the density structure used by \ac{hdbscan} depends on local core distances and mutual-reachability distances. Increasing the local multiplicity around a high-confidence rogue sample reduces its effective core distance and consequently strengthens the corresponding dense region. Hence, samples that strongly deviate from the registered-device distribution contribute more strongly to the formation of persistent novel-device clusters, whereas borderline rejections remain more likely to be treated as noise. This is precisely the information that would be lost if \ac{osr} were reduced to the binary decision \eqref{eqn:rogue_decision} before performing \ac{ncd}.

We apply \ac{hdbscan} to $\widetilde{\mySet{Z}}_{\rm R}$, yielding an estimated number $\widehat{K}_{\rm new}$ of novel devices. The cluster labels of the replicated points are projected back to each original sample by majority voting over its non-noise replicas; an original sample whose replicas are all labeled as noise remains unassigned. Denoting the resulting label of $\myVec{z}^{(j)}$ by $\widehat{c}^{(j)}$, the latent samples associated with discovered device $\ell$ are
\begin{equation}
\mySet{Z}_{\ell} = \left\{ \myVec{z}^{(j)} : \widehat{c}^{(j)}=\ell \right\},
\qquad
\ell=1,\ldots,\widehat{K}_{\rm new}.
\label{eqn:discovered_cluster}
\end{equation}

Finally, the newly discovered devices are incorporated into the same representation used for closed-set identification. For every $\mySet{Z}_{\ell}$, the \ac{lbg} algorithm is applied to obtain a codebook $\mySet{Q}_{\ell}^{\star}$, which is assigned a new device identifier $i_{\ell}^{\rm new}$. The registered-device set and codebook are then expanded according to
\begin{align}
\mySet{I} &\leftarrow \mySet{I} \cup
\left\{ i_{1}^{\rm new}, \ldots,
i_{\widehat{K}_{\rm new}}^{\rm new} \right\},
 \quad 
\mySet{Q} \leftarrow \mySet{Q}
\cup
\bigcup_{\ell=1}^{\widehat{K}_{\rm new}} \mySet{Q}_{\ell}^{\star}.
\label{eqn:device_set_expansion}
\end{align}
In addition, the newly registered samples are incorporated into the registered latent set $\mySet{Z}_{\mySet{I}}$, and the \ac{osr} model of Subsection~\ref{subsec:rogue_detection} is updated accordingly. Thus, discovery feeds back into both subsequent closed-set identification and \ac{osr}, while the feature extractor $f_{\myVec{\theta}}(\cdot)$ remains unchanged.

\subsubsection{Algorithm Summary}
\label{subsec:overall_algorithm}

The complete procedure is summarized in Algorithm~\ref{alg:holistic_rffi}. During regular operation, each received signal is mapped into the latent space and assigned an open-set score. Samples consistent with the registered set are identified through the labeled codebook, whereas rejected samples and their continuous \ac{osr} scores are accumulated for discovery. Consequently, \ac{osr} serves two purposes: it determines which samples should not be assigned to an existing device, and it provides side information regarding the reliability of these rejected samples for \ac{ncd}. Once a discovery episode is triggered, \ac{osrwc} transfers this information to \ac{hdbscan}; the resulting novel classes are then used to expand the codebook and the registered-device density model. The algorithm thus forms a closed adaptive loop in which information flows from identification to \ac{osr}, from \ac{osr} to \ac{ncd}, and from \ac{ncd} back to future identification and open-set decisions.

\begin{algorithm}[t]
\caption{Holistic \ac{rffi}}
\label{alg:holistic_rffi}

\KwData{
Feature extractor $f_{\myVec{\theta}}$;
registered  set $\mySet{I}$;
 codebook $\mySet{Q}$;\\
registered latent set $\mySet{Z}_{\mySet{I}}$;
\ac{osr} model $(\mySet{C}_{\mySet{I}},\tau)$;
\ac{osrwc} parameters
${\gamma_{\ell},\rho_{\ell}}_{\ell=1}^{L}, \sigma_{\rm jit}$;
\ac{hdbscan} parameters.
}

Initialize discovery buffer $\mySet{B}\leftarrow\emptyset$;

\ForEach{received signal ${y(t)}_{t=1}^{T}$}{
Extract
$\myVec{z} \leftarrow f_{\myVec{\theta}} \left( {y(t)}_{t=1}^{T} \right)$;
 
Compute
$r \leftarrow r_{\mySet{I}}(\myVec{z})$ using \eqref{eqn:osr_core_distance}\;

\eIf{$r\leq\tau$}{
    Find   $k^\star \leftarrow
    \arg\min_k  \|\myVec{z}-\myVec{q}_k\|_2^2$\;

    Identify device as     $\operatorname{Label}(k^\star)$\;
}{
    Declare device as  unknown\;

    Store  latent and uncertainty: $\mySet{B}   \leftarrow \mySet{B}\cup\{(\myVec{z},r)\}$\;
}

\If{the discovery criterion is satisfied}{
    Construct     $\widetilde{\mySet{Z}}_{\rm R}$   from $\mySet{B}$ using     \eqref{eqn:osrwc_weight}-\eqref{eqn:osrwc_surrogate_set}\;

    Cluster     $\widetilde{\mySet{Z}}_{\rm R}$  using \ac{hdbscan} and project  labels\;

    \For{$\ell=1,\ldots,\widehat{K}_{\rm new}$}{
        Form discovered cluster $\mySet{Z}_{\ell}$\;

        Obtain  $\mySet{Q}_{\ell}^{\star}  \leftarrow \operatorname{LBG}(\mySet{Z}_{\ell})$\;

        Assign a new device label $i_{\ell}^{\rm new}$\;
    }

    Expand $\mySet{I}$ and $\mySet{Q}$ using \eqref{eqn:device_set_expansion}\;

    Add  discovered samples to  $\mySet{Z}_{\mySet{I}}$\;
    
    Update $(\mySet{C}_{\mySet{I}},\tau)$ via  \ac{osr} fitting (Subsec.~\ref{subsec:rogue_detection})\;

    $\mySet{B}\leftarrow\emptyset$\;
} 

}
\end{algorithm}

\subsection{Training Algorithm}
\label{subsec:rffi_training}
Our training procedure separates the learning of a reusable RF representation from the registration of the devices encountered in a particular deployment. It therefore consists of two phases. First, we {\em pre-train} the feature extractor to produce a device-discriminative latent representation that is also compatible with vector quantization. Then, during {\em enrollment}, the feature extractor is frozen and the codebook and \ac{osr} model are fitted to the actual registered-device set. This separation is particularly useful in the open-world setting considered here, since changing the registered set does not require retraining the feature extractor.
To distinguish these phases, let $\mySet{D}_{\rm pre}$ denote the labeled dataset used for representation pretraining, with corresponding device set $\mySet{I}_{\rm pre}$, and let $\mySet{D}_{\rm en}$ denote the labeled enrollment data associated with the registered set $\mySet{I}$. The two datasets may correspond to the same devices, but this is not required.

\subsubsection{Representation Pretraining}
\label{subsec:representation_pretraining}
Pretraining is carried out in four successive stages. The stages gradually transition from learning a discriminative continuous representation to optimizing this representation for its eventual vector-quantized use.

\smallskip
\noindent
{\bf Stage 1 (Contrastive representation learning):}
We first train the feature extractor $f_{\myVec{\theta}}(\cdot)$ using supervised contrastive learning~\cite{khosla2020supervised}. For each training signal, two augmented views are generated using RF-domain transformations, including random phase rotation, circular time shifting, additive noise, and amplitude scaling. The latter captures variations in received signal strength due to the propagation channel.

A temporary projection head $p_{\myVec{\xi}}(\cdot)$ is applied to the resulting latent vectors, and the standard supervised contrastive objective is minimized such that representations originating from the same transmitter are attracted while those associated with different transmitters are separated. The projection head is discarded after this stage. Consequently, the output of Stage~1 is a feature extractor whose continuous latent space exhibits compact device-specific regions and separation between different devices.

\smallskip
\noindent
{\bf Stage 2 (Classifier fitting):}
We next freeze $\myVec{\theta}$ and train an auxiliary classifier $g_{\myVec{\phi}}(\cdot)$ over the continuous latent representation using the cross-entropy objective
\begin{equation}
\mathcal{L}_{\rm CE} = \frac{1}{|\mySet{D}_{\rm pre}|} \sum_{j=1}^{|\mySet{D}_{\rm pre}|} \ell_{\rm CE} \left( g_{\myVec{\phi}} \left( f_{\myVec{\theta}} \left( {y^{(j)}(t)}_{t=1}^{T} \right) \right), i^{(j)} \right).
\label{eqn:classifier_ce_loss}
\end{equation}
This stage provides a task-oriented classifier for the latent representation while preserving the geometry obtained through contrastive learning. The classifier is auxiliary: its role is to guide the subsequent quantization-aware optimization rather than to constitute the final identification rule.

\smallskip
\noindent
{\bf Stage 3: Codebook initialization}
We next initialize an auxiliary pretraining codebook $\mySet{Q}_{\rm pre}$ from the continuous latent representations of $\mySet{D}_{\rm pre}$. Specifically, the frozen feature extractor is applied to all pretraining samples, and the resulting latent vectors are quantized using the \ac{lbg} algorithm reviewed in Subsection~\ref{subsec:vector_quantization}. Initializing the codewords from the learned latent distribution avoids jointly optimizing the feature extractor and an initially unstructured codebook.

\smallskip
\noindent
{\bf Stage 4: Quantization-aware refinement}
Finally, the feature extractor, auxiliary codebook, and classifier are jointly refined. For a latent representation $\myVec{z}=f_{\myVec{\theta}}(\{y(t)\}_{t=1}^{T})$, let $\myVec{q}$ denote its nearest codeword in $\mySet{Q}_{\rm pre}$. We employ the \ac{vqvae}-based objective introduced in \eqref{eqn:train_loss}, with the task-dependent term instantiated by device-classification cross entropy:
\begin{align}
\mathcal{L}_{\rm pre} =
&\ell_{\rm CE} \left( g_{\myVec{\phi}}(\myVec{q}),i \right) + \left\| {\rm sg}(\myVec{z})-\myVec{q} \right\|_2^2
\notag\\
&+
\beta \left\| \myVec{z}-{\rm sg}(\myVec{q}) \right\|_2^2.
\label{eqn:pretraining_loss}
\end{align}
This final stage adapts the continuous representation and the discrete codebook to one another, such that quantization preserves the device-discriminative structure established in the preceding stages. After pretraining, the auxiliary classifier and codebook need not be retained; the principal output is the feature extractor $f_{\myVec{\theta}}(\cdot)$, which has been trained to produce representations suitable for device discrimination and vector quantization.

\subsubsection{Device Enrollment}
\label{subsec:enrollment}
The pre-trained representation is adapted to the registered-device set $\mySet{I}$ through enrollment. Unlike representation pretraining, enrollment modifies only lightweight components of the system and leaves $f_{\myVec{\theta}}(\cdot)$ fixed. Consequently, the registered-device set can be changed or expanded without retraining the feature extractor.

The enrollment samples in $\mySet{D}_{\rm en}$ are first mapped into the continuous latent space, yielding
\begin{equation}
\mySet{Z}_{\mySet{I}} = \left\{ f_{\myVec{\theta}} \left( \{y^{(j)}(t)\}_{t=1}^{T} \right) \right\}_{j=1}^{|\mySet{D}_{\rm en}|}.
\label{eqn:enrollment_latent_set}
\end{equation}
The operational codebook $\mySet{Q}$ is initialized by applying \ac{lbg} to $\mySet{Z}_{\mySet{I}}$. We then introduce a temporary classifier whose output dimension is $|\mySet{I}|$ and jointly refine this classifier and the codewords using the same quantization-aware objective as in \eqref{eqn:pretraining_loss}, while keeping $\myVec{\theta}$ fixed. Thus, the codebook is adapted to the enrolled devices without altering the common latent representation.

Once refinement is completed, each codeword is associated with a registered device according to
\begin{equation}
\operatorname{Label}(k) = \mathop{\arg\max}_{i\in\mySet{I}} \left[ g_{\myVec{\phi}} \left( \myVec{q}_k \right) \right]_i,
\qquad
k=1,\ldots,|\mySet{Q}|.
\label{eqn:codeword_labeling}
\end{equation}
The temporary classifier is then discarded, such that subsequent closed-set identification relies solely on the  rule in \eqref{eqn:codebook_classification}.
Finally, the  enrollment representations $\mySet{Z}_{\mySet{I}}$ are used to fit the density-based \ac{osr} model of Subsection~\ref{subsec:rogue_detection}. This yields the registered-device core set $\mySet{C}_{\mySet{I}}$ and the rejection threshold $\tau$. Enrollment therefore produces all deployment-specific quantities required by Algorithm~\ref{alg:holistic_rffi}: the labeled codebook $\mySet{Q}$, the registered latent set $\mySet{Z}_{\mySet{I}}$, and the \ac{osr} model $(\mySet{C}_{\mySet{I}},\tau)$.

\subsubsection{Training Summary}
\label{subsec:training_summary}

The complete training and enrollment procedure is summarized in Algorithm~\ref{alg:rffi_training}. The multi-stage construction serves two complementary purposes. Contrastive learning first establishes a discriminative continuous geometry, while quantization-aware refinement adapts this geometry to the discrete representation required by the proposed identification mechanism. Enrollment then decouples this representation from the particular registered-device set by fitting only the codebook and \ac{osr} model to the deployment-specific devices.

\begin{algorithm}
\caption{Training and Enrollment}
\label{alg:rffi_training}

\KwData{
Pretraining and enroll $\mySet{D}_{\rm pre}, \mySet{D}_{\rm en}$;
codebook size $P$.
}

\nonl
\textbf{Representation pretraining:} 

Train $f_{\myVec{\theta}}$ and 
$p_{\myVec{\xi}}$ via  contrastive learning on
$\mySet{D}_{\rm pre}$;


Train auxiliary classifier $g_{\myVec{\phi}}$ using
\eqref{eqn:classifier_ce_loss};

Apply \ac{lbg} to  latent representations to initialize
$\mySet{Q}_{\rm pre}$;

Jointly refine
$f_{\myVec{\theta}}$, $\mySet{Q}_{\rm pre}$, and $g_{\myVec{\phi}}$ using \eqref{eqn:pretraining_loss};

\nonl
\textbf{Device enrollment:}

Freeze $f_{\myVec{\theta}}$ and extract $\mySet{Z}_{\mySet{I}}$ from $\mySet{D}_{\rm en}$;

Initialize $\mySet{Q}\leftarrow \operatorname{LBG}(\mySet{Z}_{\mySet{I}})$;

Jointly refine $\mySet{Q}$ and a temporary classifier using \eqref{eqn:pretraining_loss};

Assign each codeword a device label using \eqref{eqn:codeword_labeling};

Fit the \ac{osr} model $(\mySet{C}_{\mySet{I}},\tau)$ from $\mySet{Z}_{\mySet{I}}$;

\Return{
$f_{\myVec{\theta}}$, $\mySet{Q}$, $\mySet{Z}_{\mySet{I}}$, $(\mySet{C}_{\mySet{I}},\tau)$
};
\end{algorithm}

\subsection{Discussion}
\label{subsec:discussion} 

\smallskip
\noindent
{\bf Holistic Open-World RFFI:}
Our framework is tailored to  open-world \ac{rffi}  by coupling the  identification tasks through a shared and expandable representation. Casting closed-set identification as vector quantization allows registered devices to be represented by  codewords, with newly discovered devices  incorporated by augmenting the codebook. The enrollment procedure further decouples representation learning from the registered-device population, enabling the same latent representation to be reused as the device set evolves. Beyond this structural coupling, our approach  transfers information between tasks: the continuous \ac{osr} score is retained after the rejection decision and exploited by \ac{osrwc} to guide \ac{ncd}, assigning greater influence to samples that more strongly deviate from the registered-device distribution. Finally, the density-based formulation is well suited to open-world operation, since the \ac{osr} model is fitted using registered-device data, while \ac{hdbscan} does not require prior knowledge of the number of unseen devices and can leave insufficiently supported samples unassigned. Our framework thus addresses the key challenges motivating our holistic formulation: it supports expansion of the registered-device set and exploits information produced by one \ac{rffi} task to improve the subsequent task.

\smallskip
\noindent
{\bf Computational Complexity:}
The computational cost is dominated by the neural feature extractor. For an input frame of $256$ IQ samples (as in Section~\ref{sec:experiments}), the complete pipeline requires approximately $180.5$ MFLOPs per frame, of which more than $99.9\%$ are associated with the convolutional feature extractor. The \ac{hcf} branch, feature fusion, codeword comparison, and density-based \ac{osr} introduce only a small additional cost. In particular, increasing the codebook size from $P=32$ to $P=256$ increases the complexity only from $180.46$ MFLOPs to $180.51$ MFLOPs per frame. Thus, expanding the registered devices through codebook augmentation has negligible impact on online inference complexity and does not require additional \ac{cnn} inference or retraining. The more involved \ac{hdbscan} and \ac{osrwc} operations are executed only when a discovery is triggered and are thus outside the continuous per-frame processing path; in the evaluated WiSig configuration, \ac{osrwc} incurs approximately $87.4$ MFLOPs per discovery episode.

\smallskip
\noindent
{\bf Possible Extensions:} 
The proposed design admits several natural extensions. The vector-quantized representation can incorporate alternative codebook-learning and quantization-aware training strategies, while the density-based \ac{osr} stage can be generalized to alternative distance or density measures better suited to particular channel conditions. Likewise, the proposed transfer of \ac{osr} information to \ac{ncd} through \ac{osrwc} represents one mechanism for coupling the two tasks; future work may consider clustering methods that incorporate the continuous open-set confidence directly. Finally, continual adaptation of the feature extractor could complement the present codebook-level expansion when long-term changes in propagation conditions or device characteristics substantially alter the learned latent representation.

\section{Experimental Study}
\label{sec:experiments}
We empirically evaluate our  \ac{rffi} framework on two RF datasets\footnote{The  code and  hyperparameters used can be found at \url{https://github.com/omer-hazan/RFFI-with-GCD-via-VQVAE}
}. Our study is designed to assess whetter: 
$(i)$  the learned vector-quantized representation preserves accurate identification of registered devices; 
$(ii)$ our density-based \ac{osr} mechanism reliably detects unseen transmitters; 
$(iii)$ transferring the \ac{osr} information to the discovery stage improves \ac{ncd}; 
and $(iv)$  newly discovered devices can be incorporated while preserving  identification and open-set capabilities. We additionally evaluate robustness to signal degradation and isolate the contribution of the proposed cross-task information transfer through an ablation study.

\subsection{Experimental Setup}

\subsubsection{Datasets}
\label{subsec:datasets}
We distinguish between the data used for representation pretraining and those used for  enrollment. Let $\mySet{I}_{\rm pre}$ denote the devices used to train the feature extractor, and  $\mySet{I}$ be the subsequently enrolled device set used to construct the  codebook and fit the registered-device \ac{osr} model. The two sets need not coincide. We further denote by $\mySet{I}_{\rm R}$ a set of unseen devices used for open-set evaluation, and by $\mySet{I}_{\rm D}\subseteq\mySet{I}_{\rm R}$ the devices presented to the discovery stage. Samples used for discovery are separated from those used to evaluate the system after registration.

\smallskip
\noindent
{\bf LoRa:}
We use the LoRa \ac{rffi} dataset~\cite{shen2021lora}, following the enrollment and rogue-device evaluation protocol of~\cite{shen2022towards} where applicable. The feature extractor is pre-trained using $30$ devices, indexed $0$--$29$, with $1000$ packets  from each device; these packets are divided into $90\%$ training and $10\%$ validation subsets. The enrolled set contains $10$ disjoint devices, indexed $30$--$39$, while devices $40$--$44$ constitute the unseen-device pool. This setting  evaluates the ability to reuse a representation learned from one transmitter population for a  different registered-device population.
For each enrolled device, packets $0$--$99$ are used for enrollment, while packets $100$--$299$ are retained for closed-set and \ac{osr} evaluation. For each unseen device, packets $0$--$99$ are used for \ac{osr} and discovery, whereas packets $100$--$199$ are held out for evaluating the system after device registration.

\smallskip
\noindent
{\bf WiSig:}
We additionally use the ManyTx configuration of the WiSig dataset~\cite{hanna2022wisig}. The setting follows the anomaly-recognition protocol of~\cite{zhao2024anomaly} where applicable and extends it to \ac{ncd} and registration. The feature extractor is pre-trained using $50$ transmitters, with $72$ packets selected from each device and divided into $80\%$, $10\%$, and $10\%$ subsets for training, validation, and testing, respectively. Ten of these devices are subsequently enrolled, such that, unlike the LoRa setting, the registered devices have already been observed during representation pretraining.
For each enrolled transmitter, $70\%$ of the available packets are used for codebook construction, quantization-aware enrollment, and fitting the registered-device density model, while the remaining $30\%$ are retained for evaluation. The unseen transmitters are similarly divided between samples used for \ac{osr}/discovery and held-out samples used after registration.

\subsubsection{Compared Algorithms}
We compare the proposed framework with five representative approaches: 
$(i)$ The method of Shen et al.~\cite{shen2022towards} (which supports only closed-set identification and \ac{osr} and is thus not used for evaluating \ac{ncd}); 
$(ii)$ The ART framework of~\cite{zhao2024anomaly}; 
$(iii)$ The GCD algorithm of~\cite{vaze2022gcd};
$(iv)$ OpenGCD~\cite{gao2023opengcdassistingopenworld}; 
and $(v)$ OpenNCD~\cite{liu2023openworldsemisupervisednovelclass}. 
While $(i)-(ii)$ are designed specifically for RFFI, benchmarks $(iii)-(v)$ were originally designed for image data, and were retrained for RF signals. 

\subsubsection{Performance Measures}

Closed-set identification is evaluated using classification accuracy. For \ac{osr}, we report the \ac{auc}, which evaluates discrimination independently of a particular operating threshold. At the calibrated threshold, we additionally report the numbers of true positives, false positives, true negatives, and false negatives, where unseen-device samples constitute the positive class.
\ac{ncd} is evaluated using the percentage of held-out samples from discovered devices that are correctly associated with their  registered identity, referred to as {\em new-device accuracy}. We also examine the effect of system expansion by comparing closed-set identification accuracy before and after newly discovered devices are incorporated. Where applicable, clustering quality is quantified using the \ac{ari}, while post-discovery \ac{osr} can be evaluated against devices in $\mySet{I}_{\rm R}\setminus\mySet{I}_{\rm D}$ that remain unseen after the discovery episode.



\begin{figure}[t]
	\centering
	\begin{subfigure}[t]{0.23\columnwidth}
		\centering
		\includegraphics[width=\linewidth]{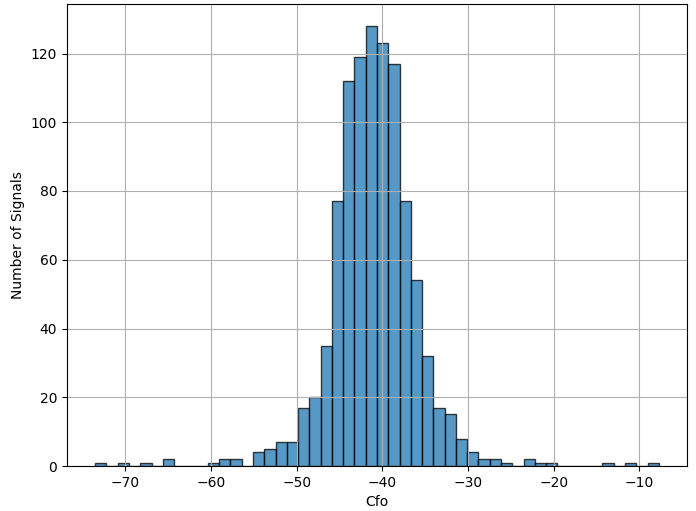}
		\caption{CFO - Device A}
	\end{subfigure}
	\hfill
	\begin{subfigure}[t]{0.23\columnwidth}
		\centering
		\includegraphics[width=\linewidth]{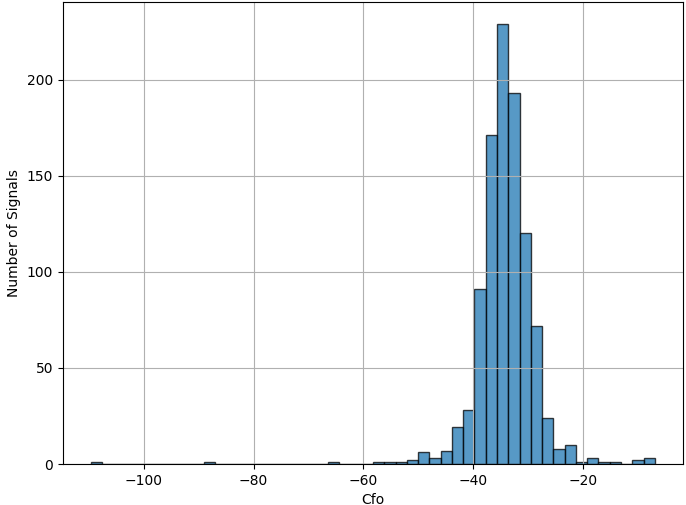}
		\caption{CFO - Device B}
	\end{subfigure}
	\hfill
%
	\begin{subfigure}[t]{0.23\columnwidth}
		\centering
		\includegraphics[width=\linewidth]{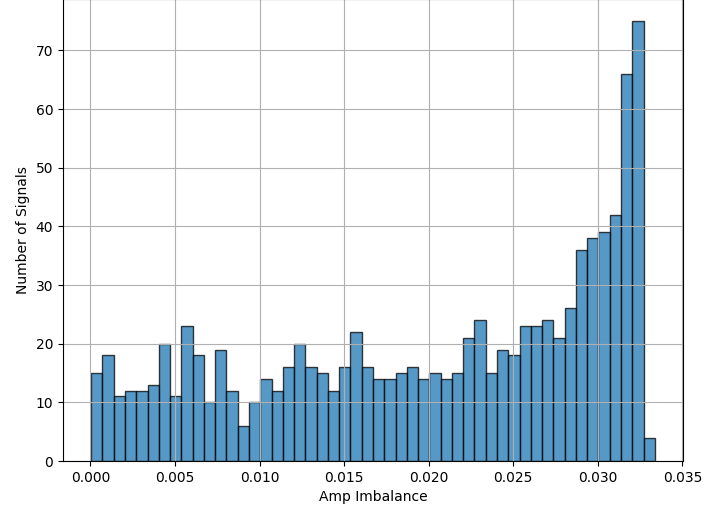}
		\caption{Amp. imbalance - Device A}
	\end{subfigure}
	\hfill
	\begin{subfigure}[t]{0.23\columnwidth}
		\centering
		\includegraphics[width=\linewidth]{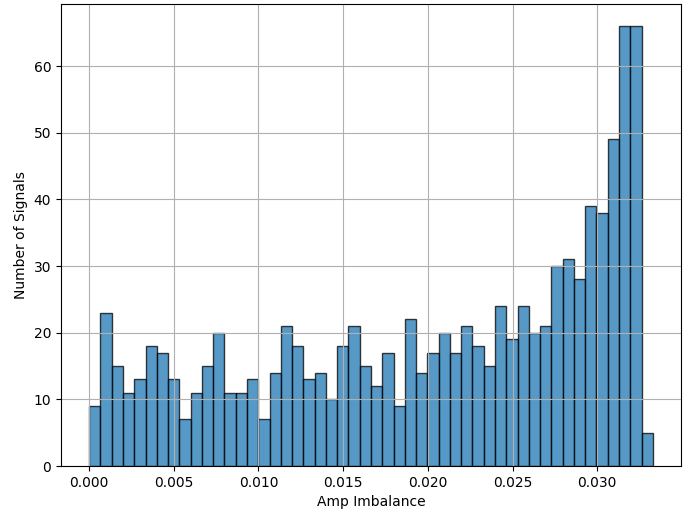}
		\caption{Amp. imbalance - Device B}
	\end{subfigure}

	\vspace{0.35cm}

	\begin{subfigure}[t]{0.23\columnwidth}
		\centering
		\includegraphics[width=\linewidth]{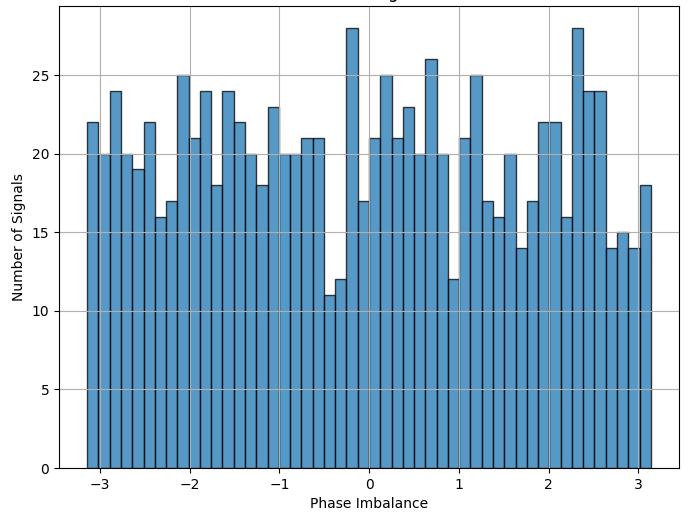}
		\caption{Phase imbalance - Device A}
	\end{subfigure}
	\hfill
	\begin{subfigure}[t]{0.23\columnwidth}
		\centering
		\includegraphics[width=\linewidth]{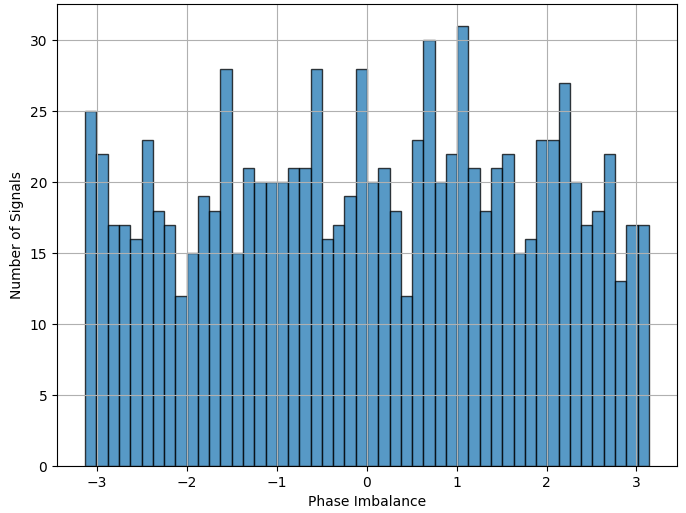}
		\caption{Phase imbalance - Device B}
	\end{subfigure}
	\hfill
	\begin{subfigure}[t]{0.23\columnwidth}
		\centering
		\includegraphics[width=\linewidth]{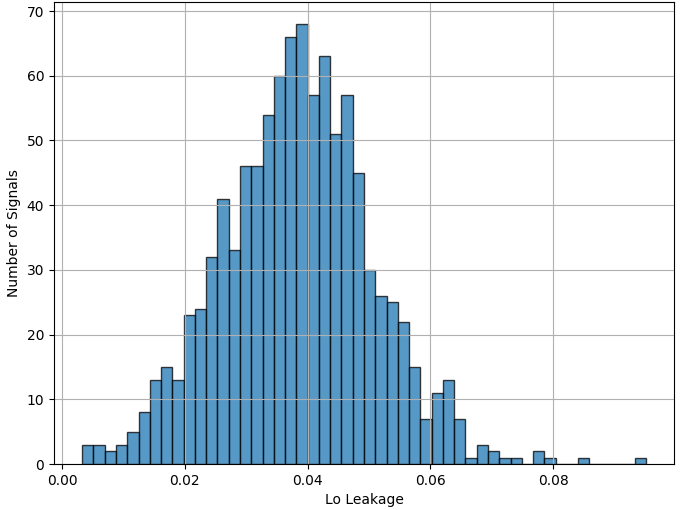}
		\caption{LO leakage - Device A}
	\end{subfigure}
	\hfill
	\begin{subfigure}[t]{0.23\columnwidth}
		\centering
		\includegraphics[width=\linewidth]{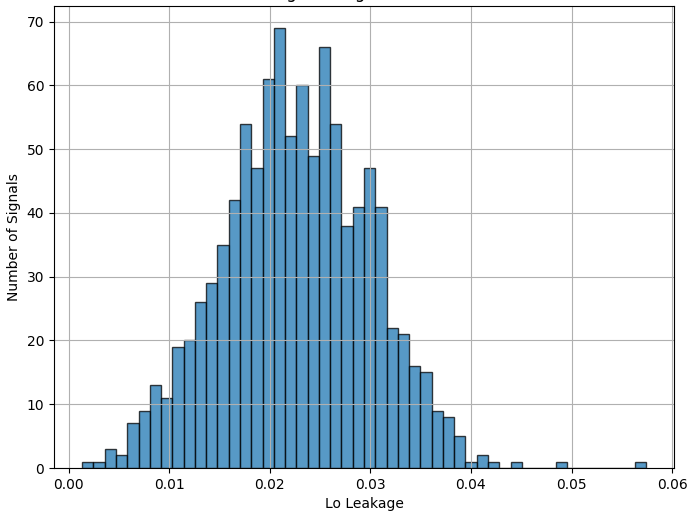}
		\caption{LO leakage - Device B}
	\end{subfigure}

	\vspace{0.5cm}

	\caption{Histogram comparison  for four  \acp{hcf}. Each row compares two devices (A and B) for a specific parameter.  
    }
	\label{fig:rf_histograms}
\end{figure}

\subsection{LoRa Evaluation}
\label{subsec:lora_evaluation}

We first consider the LoRa dataset. Unless otherwise stated, the  setting contains ten enrolled and five unseen devices.

\subsubsection{\ac{hcf} Evaluation}
\label{subsec:lora_hcf}
We begin by examining four hardware-related signal features considered as candidate inputs to the \ac{hcf} branch:  \ac{cfo}, amplitude imbalance, phase imbalance, and local-oscillator leakage. Their empirical distributions indicate that \ac{cfo}, amplitude imbalance, and local-oscillator leakage exhibit transmitter-dependent behavior, whereas phase imbalance shows substantial overlap between devices. We therefore retain the former three features in the subsequent experiments. Representative examples are shown in Fig.~\ref{fig:rf_histograms}, where device-specific distributions are clearly visible for \ac{cfo} in contrast to phase imbalance.

\begin{figure}
    \centering
    \begin{tikzpicture}
        \begin{axis}[
            width=\columnwidth,
            height=\figHeight,
            xlabel={Number of enrolled devices},
            ylabel={Accuracy (\%)},
            xmin=1,
            xmax=10,
            ymin=88,
            ymax=100.2,
            xtick={2,3,4,5,6,7,8,9,10},
            grid=major,
            legend style={
                font=\footnotesize,
                at={(0.25,0.28)},
                anchor=north,
                legend columns=2
            },
            tick label style={font=\footnotesize},
            label style={font=\footnotesize}
        ]

        \addplot+[mark=*] coordinates {
            (2,100.0)
            (3,100.0)
            (4,100.0)
            (5,100.0)
            (6,100.0)
            (7,99.89)
            (8,99.3)
            (9,98.0)
            (10,97.9)
        };
        \addlegendentry{Proposed}

        \addplot+[mark=square*] coordinates {
            (2,100.0)
            (3,100.0)
            (4,100.0)
            (5,100.0)
            (6,100.0)
            (7,100.0)
            (8,99.9)
            (9,99.9)
            (10,99.7)
        };
        \addlegendentry{ART}

        \addplot+[mark=triangle*] coordinates {
            (2,100.0)
            (3,100.0)
            (4,100.0)
            (5,100.0)
            (6,100.0)
            (7,98.29)
            (8,98.5)
            (9,97.89)
            (10,97.9)
        };
        \addlegendentry{Shen et al.}

        \addplot+[mark=diamond*] coordinates {
            (2,100.00)
            (3,100.00)
            (4,98.75)
            (5,91.00)
            (6,88.67)
            (7,92.29)
            (8,99.00)
            (9,98.67)
            (10,99.90)
        };
        \addlegendentry{GCD}

        \addplot+[color=orange, mark=pentagon] coordinates {
            (2,100.00)
            (3,100.00)
            (4,100.00)
            (5,98.80)
            (6,98.50)
            (7,98.14)
            (8,98.12)
            (9,96.89)
            (10,97.20)
        };
        \addlegendentry{OpenGCD}

        \addplot+[mark=star] coordinates {
            (2,100.00)
            (3,100.00)
            (4,100.00)
            (5,100.00)
            (6,100.00)
            (7,100.00)
            (8,100.00)
            (9,100.00)
            (10,99.90)
        };
        \addlegendentry{OpenNCD}

        \end{axis}
    \end{tikzpicture}
    \caption{
        Closed-set identification accuracy vs. $|\mySet{I}|$, LoRa.
    }
    \label{fig:lora_accuracy_vs_devices}
\end{figure}
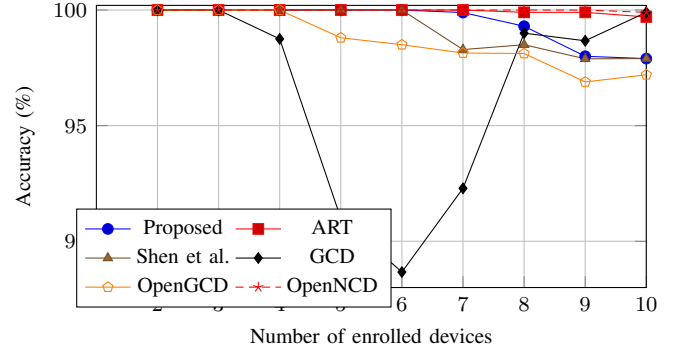

\subsubsection{Identification and \ac{osr}}
\label{subsec:lora_closed_set}
We first verify that our holistic design preserves the conventional \ac{rffi} capabilities of closed-set identification and \ac{osr}. With all ten devices enrolled, our method achieves a closed-set accuracy of $97.9\%$, matching Shen et al.~\cite{shen2022towards} and exceeding OpenGCD~\cite{gao2023opengcdassistingopenworld}, which achieves $97.2\%$. ART~\cite{zhao2024anomaly}, GCD~\cite{vaze2022gcd}, and OpenNCD~\cite{liu2023openworldsemisupervisednovelclass} achieve higher accuracies of $99.7\%$, $99.9\%$, and $99.9\%$, respectively. As shown in Fig.~\ref{fig:lora_accuracy_vs_devices}, across the enrolled-device sweep, our method maintains an accuracy between $97.9\%$ and $100\%$, consistently matching or exceeding Shen et al. 
These results indicate that the expandable codebook representation incurs only a modest cost, if any. 

\begin{figure}
    \centering
    \begin{tikzpicture}
        \begin{axis}[
            width=\columnwidth,
            height=\figHeight,
            xlabel={Number of rogue devices},
            ylabel={\ac{auc}},
            xmin=1,
            xmax=5,
            ymin=0.49,
            ymax=1.002,
            xtick={1,2,3,4,5},
            grid=major,
            tick label style={font=\footnotesize},
            label style={font=\footnotesize},
            legend style={
                font=\scriptsize,
                at={(0.97,0.45)},
                anchor=south east,
                draw=black,
                fill=white
            }
        ]

        \addplot+[mark=*] coordinates {
            (1,0.9950)
            (2,0.9967)
            (3,0.9962)
            (4,0.9934)
            (5,0.9947)
        };
        \addlegendentry{Proposed}

        \addplot+[mark=square*] coordinates {
            (1,0.9915)
            (2,0.9958)
            (3,0.9759)
            (4,0.9413)
            (5,0.9531)
        };
        \addlegendentry{Shen et al.}

        \addplot+[mark=triangle*] coordinates {
            (1,0.9438)
            (2,0.9503)
            (3,0.9454)
            (4,0.9403)
            (5,0.9435)
        };
        \addlegendentry{ART}

        \addplot+[mark=diamond*] coordinates {
            (1,0.9947)
            (2,0.9641)
            (3,0.9320)
            (4,0.9234)
            (5,0.9417)
        };
        \addlegendentry{GCD}

        \addplot+[color=orange, mark=pentagon] coordinates {
            (1,0.9156)
            (2,0.9534)
            (3,0.9234)
            (4,0.9043)
            (5,0.9214)
        };
        \addlegendentry{OpenGCD}

        \addplot+[mark=star] coordinates {
            (1,1.0000)
            (2,0.5000)
            (3,0.5000)
            (4,0.8191)
            (5,0.7948)
        };
        \addlegendentry{OpenNCD}

        \end{axis}
    \end{tikzpicture}
    \caption{
        \ac{osr} \ac{auc} vs. $|\mySet{I}_{\rm R}|$, LoRa.
    }
    \label{fig:lora_auc_vs_rogues}
\end{figure}
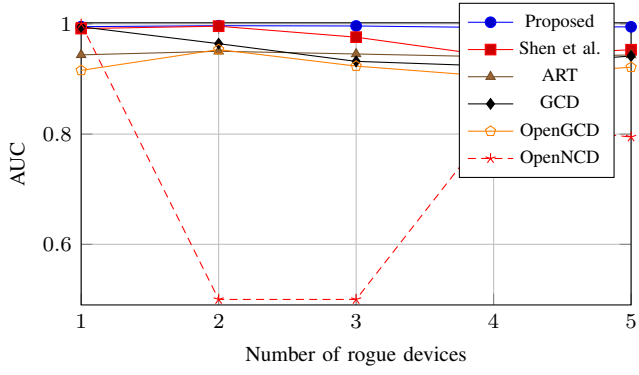

We next consider \ac{osr} with ten enrolled and five unseen devices. As reported in Table~\ref{tab:lora_osr}, our density-based score achieves the highest \ac{auc} among the evaluated methods at $0.9947$, compared with $0.9531$ for~\cite{shen2022towards}, $0.9436$ for~\cite{zhao2024anomaly}, $0.9417$ for~\cite{vaze2022gcd}, $0.8929$ for~\cite{gao2023opengcdassistingopenworld}, and $0.7948$ for~\cite{liu2023openworldsemisupervisednovelclass}. At the selected  threshold, our method correctly rejects $490$ of the $500$ unseen-device samples while accepting $959$ of the $1000$ legitimate samples. Although the methods of~\cite{vaze2022gcd} and~\cite{liu2023openworldsemisupervisednovelclass} produce no false-positive rejections at their respective operating thresholds, they detect only $145$ and $207$ unseen-device samples, respectively. Moreover, Fig.~\ref{fig:lora_auc_vs_rogues} shows that we maintain an \ac{auc} between $0.9934$ and $0.9967$ throughout the rogue-device sweep, indicating consistent separation between enrolled and unseen devices.

\begin{table}[b]
    \centering
    \caption{
        LoRa \ac{osr} with ten enrolled and five rogue
        devices. 
    }
    \label{tab:lora_osr}
    \begin{tabular}{lccccc}
        \toprule
        Method & \ac{auc} & TP & FP & TN & FN \\
        \midrule
        Proposed
            & \textbf{0.9947}
            & 490
            & \textbf{41}
            & \textbf{959}
            & 10 \\
        ART~\cite{zhao2024anomaly}
            & 0.9436
            & \textbf{496}
            & 172
            & 828
            & \textbf{4} \\
        Shen et al.~\cite{shen2022towards}
            & 0.9531
            & 436
            & 90
            & 910
            & 64 \\
        GCD~\cite{vaze2022gcd}
            & 0.9417
            & 145
            & 0
            & 1000
            & 355 \\
        OpenGCD~\cite{gao2023opengcdassistingopenworld}
            & 0.8929
            & 337
            & 70
            & 930
            & 163 \\
        OpenNCD~\cite{liu2023openworldsemisupervisednovelclass}
            & 0.7948
            & 207
            & 0
            & 1000
            & 293 \\
        \bottomrule
    \end{tabular}
\end{table}

\subsubsection{Generalized Category Discovery}
\label{subsec:lora_gcd}
\begin{figure}
    \centering
    \begin{tikzpicture}
        \begin{axis}[
            width=\columnwidth,
            height=\figHeight,
            xlabel={Number of rogue devices},
            ylabel={Mapped rogue samples (\%)},
            xmin=2,
            xmax=5,
            ymin=0,
            ymax=100,
            xtick={2,3,4,5},
            ytick={0,10,20,30,40,50,60,70,80,90,100},
            grid=major,
            tick label style={font=\footnotesize},
            label style={font=\footnotesize},
            legend style={
                font=\scriptsize,
                at={(0.97,0.97)},
                anchor=north east,
                draw=black,
                fill=white
            }
        ]

        \addplot+[mark=*] coordinates {
            (2,90.0)
            (3,79.3)
            (4,93.0)
            (5,81.8)
        };
        \addlegendentry{Proposed}

        \addplot+[mark=triangle*] coordinates {
            (2,80.5)
            (3,71.7)
            (4,56.8)
            (5,63.8)
        };
        \addlegendentry{ART}

        \addplot+[mark=diamond*] coordinates {
            (2,19.5)
            (3,98.0)
            (4,24.8)
            (5,29.4)
        };
        \addlegendentry{GCD}

        \addplot+[color=orange, mark=pentagon] coordinates {
            (2,84.0)
            (3,83.7)
            (4,80.2)
            (5,80.6)
        };
        \addlegendentry{OpenGCD}

        \addplot+[mark=star] coordinates {
            (2,0.0)
            (3,0.0)
            (4,25.0)
            (5,18.6)
        };
        \addlegendentry{OpenNCD}

        \end{axis}
    \end{tikzpicture}
    \caption{
        Percentage of held-out LoRa rogue samples mapped to newly
        discovered devices vs. $|\mySet{I}_{\rm R}|$.
    }
    \label{fig:lora_mapping_vs_rogues}
\end{figure}
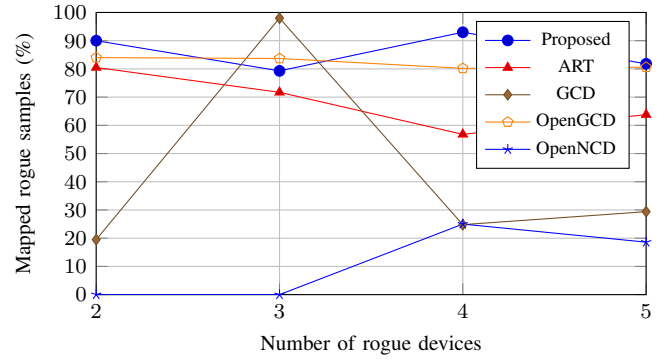

We next evaluate the complete open-world operation, where rejected signals are clustered, newly discovered devices are registered, and subsequent packets are classified using the expanded system. With five unseen devices, the proposed method correctly associates $81.8\%$ of the held-out samples with their newly registered identities, compared with $80.6\%$ for~\cite{gao2023opengcdassistingopenworld}, $57.7\%$ for~\cite{zhao2024anomaly}, $29.4\%$ for~\cite{vaze2022gcd}, and $18.6\%$ for~\cite{liu2023openworldsemisupervisednovelclass}, as summarized in Table~\ref{tab:lora_gcd}. Fig.~\ref{fig:lora_mapping_vs_rogues} shows that the proposed method leads in three of the four discovery settings, while the method of~\cite{vaze2022gcd} performs best with three unseen devices. 
We note that the benefit is not limited to the discovery stage itself. After incorporating the new devices, the proposed method achieves the highest total accuracy at $91.1\%$, compared with $86.1\%$ for the closest comparator~\cite{gao2023opengcdassistingopenworld}. Its accuracy on previously enrolled devices decreases from $97.9\%$ to $95.8\%$. Although the methods of~\cite{vaze2022gcd} and~\cite{liu2023openworldsemisupervisednovelclass} preserve higher enrolled-device accuracies, their lower new-device accuracies limit overall performance. Thus, the expandable codebook supports discovery and registration while largely preserving identification accuracy on previously enrolled devices.



\begin{table}[b]
    \centering
    \caption{
        LoRa discovery  performance, $|\mySet{{I}}_{{\rm R}}|=5$. The arrow denotes pre-discovery accuracy on the enrolled
        devices to post-discovery accuracy over all enrolled and rogue test samples;
        in brackets, post-discovery accuracy on the enrolled devices only.
    }
    \label{tab:lora_gcd}
    \begin{tabular}{lcc}
        \toprule
        Method & New-device accuracy & Total accuracy (enrolled) \\
        \midrule
        Proposed
            & \textbf{81.8\%}
            & $97.9\% \rightarrow \mathbf{91.1\%}~(95.8\%)$ \\
        ART~\cite{zhao2024anomaly}
            & 57.7\%
            & $99.8\% \rightarrow 76.3\%~(88.7\%)$ \\
        GCD~\cite{vaze2022gcd}
            & 29.4\%
            & $99.7\% \rightarrow 76.3\%~(99.7\%)$ \\
        OpenGCD~\cite{gao2023opengcdassistingopenworld}
            & 80.6\%
            & $97.2\% \rightarrow 86.1\%~(88.9\%)$ \\
        OpenNCD~\cite{liu2023openworldsemisupervisednovelclass}
            & 18.6\%
            & $\mathbf{100.0\%} \rightarrow 72.9\%~(100.0\%)$ \\
        \bottomrule
    \end{tabular}
\end{table}

\subsection{WiFi Evaluation}
\label{subsec:wifi_evaluation}

We next evaluate the framework on WiSig, a complementary setting whose conditions differ substantially from  LoRa.

\subsubsection{Identification and \ac{osr}}
\label{subsec:wifi_osr}
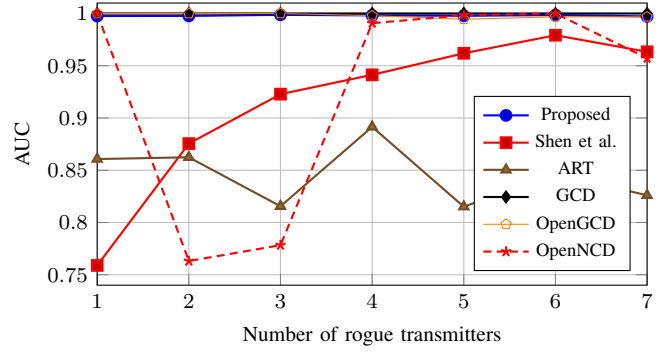
\begin{figure}
    \centering
    \begin{tikzpicture}
        \begin{axis}[
            width=\columnwidth,
            height=\figHeight,
            xlabel={Number of rogue transmitters},
            ylabel={\ac{auc}},
            xmin=1,
            xmax=7,
            ymin=0.74,
            ymax=1.01,
            xtick={1,2,3,4,5,6,7},
            ytick={0.75,0.80,0.85,0.90,0.95,1.00},
            grid=major,
            tick label style={font=\footnotesize},
            label style={font=\footnotesize},
            legend style={
                font=\scriptsize,
                at={(0.97,0.05)},
                anchor=south east,
                draw=black,
                fill=white
            }
        ]

        \addplot+[mark=*, thick] coordinates {
            (1,0.9976)
            (2,0.9977)
            (3,0.9987)
            (4,0.9982)
            (5,0.9978)
            (6,0.9985)
            (7,0.9972)
        };
        \addlegendentry{Proposed}

        \addplot+[mark=square*, thick] coordinates {
            (1,0.7589)
            (2,0.8755)
            (3,0.9228)
            (4,0.9413)
            (5,0.9619)
            (6,0.9792)
            (7,0.9633)
        };
        \addlegendentry{Shen et al.}

        \addplot+[mark=triangle*, thick] coordinates {
            (1,0.8607)
            (2,0.8624)
            (3,0.8155)
            (4,0.8915)
            (5,0.8151)
            (6,0.8467)
            (7,0.8260)
        };
        \addlegendentry{ART}

        \addplot+[mark=diamond*, thick] coordinates {
            (1,1.0000)
            (2,1.0000)
            (3,1.0000)
            (4,1.0000)
            (5,1.0000)
            (6,1.0000)
            (7,1.0000)
        };
        \addlegendentry{GCD}

        \addplot+[color=orange, mark=pentagon] coordinates {
            (1,1.0000)
            (2,1.0000)
            (3,1.0000)
            (4,0.9979)
            (5,0.9945)
            (6,0.9968)
            (7,0.9968)
        };
        \addlegendentry{OpenGCD}

        \addplot+[mark=star, thick] coordinates {
            (1,1.0000)
            (2,0.7631)
            (3,0.7783)
            (4,0.9907)
            (5,0.9984)
            (6,1.0000)
            (7,0.9572)
        };
        \addlegendentry{OpenNCD}

        \end{axis}
    \end{tikzpicture}
    \caption{
       \ac{osr} \ac{auc} vs. $|\mySet{I}_{\rm R}|$, WiSig.
    }
    \label{fig:wisig_auc_vs_rogues}
\end{figure}

The proposed method attains perfect pre-discovery closed-set accuracy in the considered WiSig setting. With ten enrolled and seven rogue transmitters, Table~\ref{tab:wifi_osr} reports an \ac{auc} of $0.9972$, compared with $1.0000$ for~\cite{vaze2022gcd} and $0.9968$ for~\cite{gao2023opengcdassistingopenworld}. The remaining methods~\cite{shen2022towards,zhao2024anomaly,liu2023openworldsemisupervisednovelclass} achieve lower \ac{auc} values between $0.8615$ and $0.9633$. At the calibrated threshold, all $175$ unseen-device samples are correctly rejected, while only nine of the $150$ legitimate samples are incorrectly rejected. Fig.~\ref{fig:wisig_auc_vs_rogues} further shows that the proposed score remains close to unit \ac{auc} throughout the rogue-device sweep.

\begin{table}[b]
    \centering
    \caption{
        WiSig \ac{osr} with $|\mySet{I}|=10$ and $|\mySet{I}_{\rm R}| = 7$.
    }
    \label{tab:wifi_osr}
    \begin{tabular}{lccccc}
        \toprule
        Method & \ac{auc} & TP & FP & TN & FN \\
        \midrule
        Proposed
            & 0.9972
            & \textbf{175}
            & 9
            & 141
            & \textbf{0} \\
        ART~\cite{zhao2024anomaly}
            & 0.8615
            & 44
            & 11
            & 139
            & 131 \\
        Shen et al.~\cite{shen2022towards}
            & 0.9633
            & 147
            & 1
            & 149
            & 28 \\
        GCD~\cite{vaze2022gcd}
            & \textbf{1.0000}
            & 160
            & \textbf{0}
            & \textbf{150}
            & 15 \\
        OpenGCD~\cite{gao2023opengcdassistingopenworld}
            & 0.9968
            & 160
            & \textbf{0}
            & \textbf{150}
            & 15 \\
        OpenNCD~\cite{liu2023openworldsemisupervisednovelclass}
            & 0.9572
            & 134
            & 2
            & 148
            & 41 \\
        \bottomrule
    \end{tabular}
\end{table}

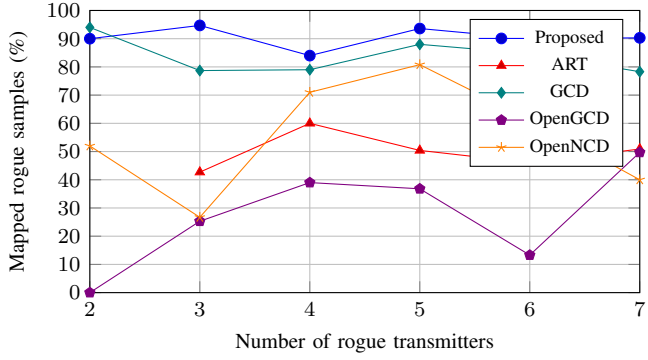
\begin{figure}
    \centering
    \begin{tikzpicture}
        \begin{axis}[
            width=\columnwidth,
            height=\figHeight,
            xlabel={Number of rogue transmitters},
            ylabel={Mapped rogue samples (\%)},
            xmin=2,
            xmax=7,
            ymin=0,
            ymax=100,
            xtick={2,3,4,5,6,7},
            ytick={0,10,20,30,40,50,60,70,80,90,100},
            grid=major,
            tick label style={font=\footnotesize},
            label style={font=\footnotesize},
            legend style={
                font=\scriptsize,
                at={(0.97,0.97)},
                anchor=north east,
                draw=black,
                fill=white
            }
        ]

        \addplot+[mark=*] coordinates {
            (2,90.0)
            (3,94.7)
            (4,84.0)
            (5,93.6)
            (6,90.0)
            (7,90.3)
        };
        \addlegendentry{Proposed}

        \addplot+[mark=triangle*] coordinates {
            (3,42.7)
            (4,60.0)
            (5,50.4)
            (6,46.0)
            (7,50.9)
        };
        \addlegendentry{ART}

        \addplot+[mark=diamond*, color=teal, mark options={solid, fill=teal}] coordinates {
            (2,94.0)
            (3,78.7)
            (4,79.0)
            (5,88.0)
            (6,84.7)
            (7,78.3)
        };
        \addlegendentry{GCD}

        \addplot+[mark=pentagon*, color=violet, mark options={solid, fill=violet}] coordinates {
            (2,0.0)
            (3,25.3)
            (4,39.0)
            (5,36.8)
            (6,13.3)
            (7,49.7)
        };
        \addlegendentry{OpenGCD}

        \addplot+[mark=star, color=orange, mark options={solid, fill=orange}] coordinates {
            (2,52.0)
            (3,26.7)
            (4,71.0)
            (5,80.8)
            (6,61.3)
            (7,40.0)
        };
        \addlegendentry{OpenNCD}

        \end{axis}
    \end{tikzpicture}
    \caption{
        Percentage of held-out WiSig rogue samples mapped to newly
        discovered transmitters vs. $|\mySet{I}_{\rm R}|$.
    }
    \label{fig:wisig_mapping_vs_rogues}
\end{figure}

\subsubsection{Generalized Category Discovery}
\label{subsec:wifi_gcd}
The WiSig results further highlight the benefit of our framework after open-set rejection. With seven rogue transmitters, the proposed method achieves a new-device accuracy of $90.3\%$, compared with $78.3\%$ for~\cite{vaze2022gcd}, $50.9\%$ for~\cite{zhao2024anomaly}, $49.7\%$ for~\cite{gao2023opengcdassistingopenworld}, and $40.0\%$ for~\cite{liu2023openworldsemisupervisednovelclass}, as summarized in Table~\ref{tab:wifi_gcd}. Fig.~\ref{fig:wisig_mapping_vs_rogues} shows that its accuracy remains between $84.0\%$ and $94.7\%$ throughout the evaluated multi-device discovery settings.
Following registration, our method achieves the highest total accuracy at $92.6\%$, compared with $88.3\%$ for the closest comparator~\cite{vaze2022gcd}. Its accuracy on previously enrolled devices decreases from $100.0\%$ to $95.3\%$. Together with the LoRa results, this shows that the quantized representation supports device-set expansion without retraining the common feature extractor. 


\begin{table}[b]
    \centering
    \caption{
        WiSig discovery performance, $|\mySet{{I}}_{{\rm R}}|=7$. The arrow denotes pre-discovery accuracy on the enrolled
        devices to post-discovery accuracy over all enrolled and rogue test samples;
        in brackets, post-discovery accuracy on the enrolled devices only.
    }
    \label{tab:wifi_gcd}
    \begin{tabular}{lcc}
        \toprule
        Method & New-device accuracy & Total accuracy (enrolled) \\
        \midrule
        Proposed
            & \textbf{90.3\%}
            & $100.0\% \rightarrow \mathbf{92.6\%}~(95.3\%)$ \\
        ART~\cite{zhao2024anomaly}
            & 50.9\%
            & $100.0\% \rightarrow 65.3\%~(82.0\%)$ \\
        GCD~\cite{vaze2022gcd}
            & 78.3\%
            & $100.0\% \rightarrow 88.3\%~(100.0\%)$ \\
        OpenGCD~\cite{gao2023opengcdassistingopenworld}
            & 49.7\%
            & $100.0\% \rightarrow 72.6\%~(99.3\%)$ \\
        OpenNCD~\cite{liu2023openworldsemisupervisednovelclass}
            & 40.0\%
            & $100.0\% \rightarrow 67.1\%~(98.7\%)$ \\
        \bottomrule
    \end{tabular}
\end{table}

\subsubsection{Robustness to Noise}
\label{subsec:wisig_awgn}
\begin{figure}
    \centering
    \begin{tikzpicture}
        \begin{axis}[
            width=\columnwidth,
            height=\figHeight,
            xlabel={SNR (dB)},
            ylabel={\ac{auc}},
            xmin=12,
            xmax=30,
            x dir=reverse,
            ymin=0.50,
            ymax=1.01,
            xtick={12,15,20,25,30},
            ytick={0.60,0.70,0.80,0.90,1.00},
            grid=major,
            tick label style={font=\footnotesize},
            label style={font=\footnotesize},
            legend style={
                font=\scriptsize,
                at={(0.97,0.05)},
                anchor=south east,
                draw=black,
                fill=white
            }
        ]

        \addplot+[
            mark=*,
            thick,
            error bars/.cd,
            y dir=both,
            y explicit
        ] coordinates {
            (30,0.9972) +- (0,0.0003)
            (25,0.9969) +- (0,0.0005)
            (20,0.9969) +- (0,0.0007)
            (15,0.9938) +- (0,0.0012)
            (12,0.9716) +- (0,0.0043)
        };
        \addlegendentry{Proposed}

        \addplot+[
            mark=triangle*,
            thick,
            error bars/.cd,
            y dir=both,
            y explicit
        ] coordinates {
            (30,0.8461) +- (0,0.0254)
            (25,0.8220) +- (0,0.0215)
            (20,0.8220) +- (0,0.0296)
            (15,0.8038) +- (0,0.0303)
            (12,0.6498) +- (0,0.0604)
        };
        \addlegendentry{ART}

        \addplot+[
            mark=square*,
            thick,
            error bars/.cd,
            y dir=both,
            y explicit
        ] coordinates {
            (30,0.9645) +- (0,0.0007)
            (25,0.9633) +- (0,0.0019)
            (20,0.9655) +- (0,0.0030)
            (15,0.9678) +- (0,0.0064)
            (12,0.9745) +- (0,0.0047)
        };
        \addlegendentry{Shen et al.}

        \addplot+[
            mark=diamond*, color=teal, mark options={solid, fill=teal},
            thick,
            error bars/.cd,
            y dir=both,
            y explicit
        ] coordinates {
            (30,0.9999) +- (0,0.0001)
            (25,0.9998) +- (0,0.0001)
            (20,0.9921) +- (0,0.0155)
            (15,0.9982) +- (0,0.0020)
            (12,0.9883) +- (0,0.0086)
        };
        \addlegendentry{GCD}

        \addplot+[
            mark=pentagon*, color=violet, mark options={solid, fill=violet},
            thick,
            error bars/.cd,
            y dir=both,
            y explicit
        ] coordinates {
            (30,0.9899) +- (0,0.0021)
            (25,0.9881) +- (0,0.0029)
            (20,0.9876) +- (0,0.0034)
            (15,0.9826) +- (0,0.0056)
            (12,0.9714) +- (0,0.0091)
        };
        \addlegendentry{OpenGCD}

        \addplot+[
            mark=star, color=orange, mark options={solid, fill=orange},
            thick,
            error bars/.cd,
            y dir=both,
            y explicit
        ] coordinates {
            (30,0.9430) +- (0,0.0343)
            (25,0.9732) +- (0,0.0198)
            (20,0.9864) +- (0,0.0145)
            (15,0.9820) +- (0,0.0281)
            (12,0.9888) +- (0,0.0078)
        };
        \addlegendentry{OpenNCD}

        \end{axis}
    \end{tikzpicture}
    \caption{
        WiSig \ac{osr} \ac{auc} vs. evaluation \ac{snr}. 
    }
    \label{fig:wisig_awgn_osr_auc}
\end{figure}
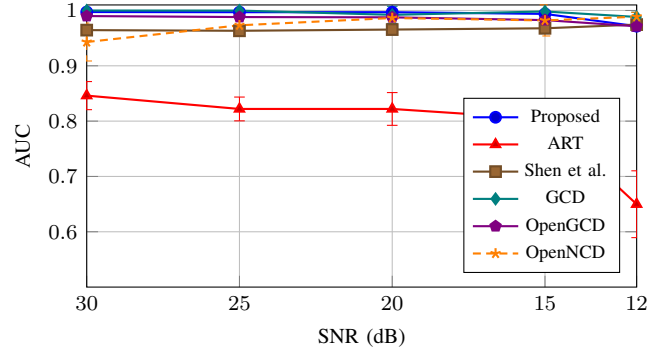

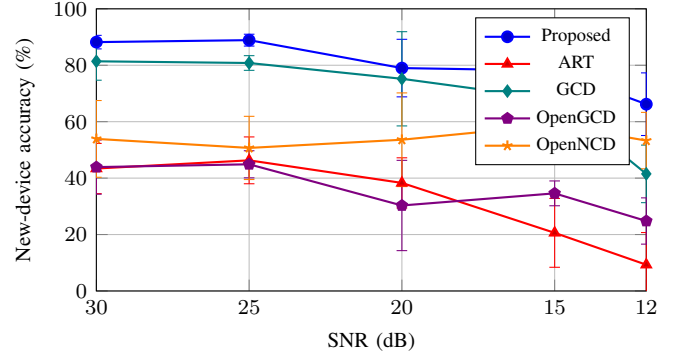
\begin{figure}
    \centering
    \begin{tikzpicture}
        \begin{axis}[
            width=\columnwidth,
            height=\figHeight,
            xlabel={SNR (dB)},
            ylabel={New-device accuracy (\%)},
            xmin=12,
            xmax=30,
            x dir=reverse,
            ymin=0,
            ymax=100,
            xtick={12,15,20,25,30},
            ytick={0,20,40,60,80,100},
            grid=major,
            tick label style={font=\footnotesize},
            label style={font=\footnotesize},
            legend style={
                font=\scriptsize,
                at={(0.97,0.97)},
                anchor=north east,
                draw=black,
                fill=white
            }
        ]

        \addplot+[
            mark=*,
            thick,
            error bars/.cd,
            y dir=both,
            y explicit
        ] coordinates {
            (30,88.2) +- (0,2.4)
            (25,88.9) +- (0,2.1)
            (20,79.0) +- (0,10.2)
            (15,78.1) +- (0,10.4)
            (12,66.2) +- (0,11.1)
        };
        \addlegendentry{Proposed}

        \addplot+[
            mark=triangle*,
            thick,
            error bars/.cd,
            y dir=both,
            y explicit
        ] coordinates {
            (30,43.4) +- (0,8.9)
            (25,46.3) +- (0,8.3)
            (20,38.3) +- (0,8.9)
            (15,20.6) +- (0,12.2)
            (12,9.3) +- (0,11.4)
        };
        \addlegendentry{ART}

        \addplot+[
            mark=diamond*, color=teal, mark options={solid, fill=teal},
            thick,
            error bars/.cd,
            y dir=both,
            y explicit
        ] coordinates {
            (30,81.4) +- (0,6.7)
            (25,80.8) +- (0,2.6)
            (20,75.2) +- (0,16.7)
            (15,67.9) +- (0,11.2)
            (12,41.5) +- (0,10.2)
        };
        \addlegendentry{GCD}

        \addplot+[
            mark=pentagon*, color=violet, mark options={solid, fill=violet},
            thick,
            error bars/.cd,
            y dir=both,
            y explicit
        ] coordinates {
            (30,43.9) +- (0,9.6)
            (25,44.9) +- (0,4.8)
            (20,30.3) +- (0,16.0)
            (15,34.6) +- (0,4.4)
            (12,24.8) +- (0,8.2)
        };
        \addlegendentry{OpenGCD}

        \addplot+[
            mark=star, color=orange, mark options={solid, fill=orange},
            thick,
            error bars/.cd,
            y dir=both,
            y explicit
        ] coordinates {
            (30,53.9) +- (0,13.6)
            (25,50.7) +- (0,11.2)
            (20,53.6) +- (0,16.6)
            (15,58.9) +- (0,12.2)
            (12,53.3) +- (0,10.0)
        };
        \addlegendentry{OpenNCD}

        \end{axis}
    \end{tikzpicture}
    \caption{
        WiSig new-device accuracy vs. evaluation \ac{snr}. 
    }
    \label{fig:wisig_awgn_new_device}
\end{figure}

We next evaluate the complete pipeline under Gaussian noise  added  to the evaluation signals (feature-extractor training and enrollment remain unchanged). We consider \ac{snr} values from $30$ to $12$ dB. As shown in Fig.~\ref{fig:wisig_awgn_osr_auc}, our method maintains an \ac{osr} \ac{auc} above $0.97$ throughout the evaluated range. Although some baselines achieve slightly higher \ac{auc}, this advantage does not translate into improved discovery. At $12$ dB, the proposed method trails~\cite{vaze2022gcd} and~\cite{liu2023openworldsemisupervisednovelclass} by $0.0167$ and $0.0172$ in \ac{auc}, respectively, yet its new-device accuracy of $66.2\%$ exceeds theirs by $24.7$ and $12.9$ percentage points. Moreover, Fig.~\ref{fig:wisig_awgn_new_device} shows that the proposed method achieves the highest mean new-device accuracy at every evaluated \ac{snr}. For the complete open-world task, this substantial accuracy advantage outweighs the modest shortfall in detection \ac{auc}.


Table~\ref{tab:wisig_awgn_closed_set} further shows that our method consistently achieves the highest total post-discovery accuracy. Even at $12$ dB \ac{snr}, it attains $77.3\%$, compared with $73.3\%$ for~\cite{liu2023openworldsemisupervisednovelclass} and $68.2\%$ for~\cite{vaze2022gcd}, while maintaining $90.3\%$ accuracy on previously enrolled devices. The discovery stage of~\cite{zhao2024anomaly} fails in this setting. Thus, our framework retains its overall identification advantage after registering newly discovered devices, even under substantial noise.
\begin{table}
    \centering
    \caption{
        WiSig discovery performance vs. \ac{snr}. Each cell:
        pre-discovery accuracy on the enrolled devices $\rightarrow$ post-discovery
        accuracy over all enrolled and rogue test samples;  in brackets,
        post-discovery accuracy on the enrolled devices only.
        \textsuperscript{\dag}\,\ac{ncd} failed.
    }
    \label{tab:wisig_awgn_closed_set}
    \footnotesize
    \setlength{\tabcolsep}{3pt}
    \resizebox{\columnwidth}{!}{
    \begin{tabular}{cccccc}
        \toprule
        \begin{tabular}[c]{@{}c@{}}\ac{snr}\\(dB)\end{tabular} & Proposed & \begin{tabular}[c]{@{}c@{}}ART\\\cite{zhao2024anomaly}\end{tabular} & \begin{tabular}[c]{@{}c@{}}GCD\\\cite{vaze2022gcd}\end{tabular} & \begin{tabular}[c]{@{}c@{}}OpenGCD\\\cite{gao2023opengcdassistingopenworld}\end{tabular} & \begin{tabular}[c]{@{}c@{}}OpenNCD\\\cite{liu2023openworldsemisupervisednovelclass}\end{tabular} \\
        \midrule
        30 & \begin{tabular}[c]{@{}c@{}}100.0$\rightarrow$\textbf{91.2}\\(94.8)\end{tabular}
           & \begin{tabular}[c]{@{}c@{}}100.0$\rightarrow$66.3\\(93.1)\end{tabular}
           & \begin{tabular}[c]{@{}c@{}}100.0$\rightarrow$90.0\\(100.0)\end{tabular}
           & \begin{tabular}[c]{@{}c@{}}100.0$\rightarrow$69.4\\(99.2)\end{tabular}
           & \begin{tabular}[c]{@{}c@{}}98.7$\rightarrow$71.8\\(92.5)\end{tabular} \\[2pt]
        25 & \begin{tabular}[c]{@{}c@{}}100.0$\rightarrow$\textbf{91.8}\\(95.1)\end{tabular}
           & \begin{tabular}[c]{@{}c@{}}100.0$\rightarrow$68.3\\(93.9)\end{tabular}
           & \begin{tabular}[c]{@{}c@{}}100.0$\rightarrow$89.7\\(100.0)\end{tabular}
           & \begin{tabular}[c]{@{}c@{}}100.0$\rightarrow$70.1\\(99.5)\end{tabular}
           & \begin{tabular}[c]{@{}c@{}}100.0$\rightarrow$71.3\\(95.3)\end{tabular} \\[2pt]
        20 & \begin{tabular}[c]{@{}c@{}}100.0$\rightarrow$\textbf{86.9}\\(96.1)\end{tabular}
           & \begin{tabular}[c]{@{}c@{}}100.0$\rightarrow$62.3\\(90.3)\end{tabular}
           & \begin{tabular}[c]{@{}c@{}}100.0$\rightarrow$86.6\\(100.0)\end{tabular}
           & \begin{tabular}[c]{@{}c@{}}99.9$\rightarrow$57.7\\(98.7)\end{tabular}
           & \begin{tabular}[c]{@{}c@{}}99.7$\rightarrow$70.2\\(89.5)\end{tabular} \\[2pt]
        15 & \begin{tabular}[c]{@{}c@{}}98.0$\rightarrow$\textbf{85.6}\\(94.3)\end{tabular}
           & \begin{tabular}[c]{@{}c@{}}99.9$\rightarrow$52.3\\(89.2)\end{tabular}
           & \begin{tabular}[c]{@{}c@{}}100.0$\rightarrow$82.7\\(100.0)\end{tabular}
           & \begin{tabular}[c]{@{}c@{}}99.6$\rightarrow$63.6\\(97.5)\end{tabular}
           & \begin{tabular}[c]{@{}c@{}}98.5$\rightarrow$71.9\\(87.2)\end{tabular} \\[2pt]
        12 & \begin{tabular}[c]{@{}c@{}}93.3$\rightarrow$\textbf{77.3}\\(90.3)\end{tabular}
           & \begin{tabular}[c]{@{}c@{}}99.1$\rightarrow$\textsuperscript{\dag}\\~\end{tabular}
           & \begin{tabular}[c]{@{}c@{}}99.9$\rightarrow$68.2\\(99.5)\end{tabular}
           & \begin{tabular}[c]{@{}c@{}}99.2$\rightarrow$55.9\\(92.1)\end{tabular}
           & \begin{tabular}[c]{@{}c@{}}99.7$\rightarrow$73.3\\(96.7)\end{tabular} \\[2pt]
        \bottomrule
    \end{tabular}%
    }
\end{table}

\begin{figure}
    \centering

    \subfloat[LoRa]{
        \label{fig:new_device_accuracy_lora}
        \begin{tikzpicture}
            \begin{axis}[
                width=0.95\columnwidth,
                height=0.52\columnwidth,
                xlabel={Number of rogue devices},
                ylabel={New-device accuracy (\%)},
                xmin=2,
                xmax=5,
                ymin=78,
                ymax=95,
                xtick={2,3,4,5},
                ytick={80,85,90,95},
                grid=major,
                tick label style={font=\footnotesize},
                label style={font=\footnotesize},
                legend style={
                    font=\scriptsize,
                    at={(0.37,0.973)},
                    anchor=north,
                    legend columns=2,
                    draw=black,
                    fill=white
                }
            ]

            \addplot+[mark=*, thick] coordinates {
                (2,90.0)
                (3,79.3)
                (4,93.0)
                (5,81.8)
            };
            \addlegendentry{With OSRWC}

            \addplot+[mark=square*, dashed, thick] coordinates {
                (2,89.5)
                (3,79.3)
                (4,82.8)
                (5,79.2)
            };
            \addlegendentry{Without OSRWC}

            \end{axis}
        \end{tikzpicture}
    }

    \vspace{0.25cm}

    \subfloat[WiSig]{
        \label{fig:new_device_accuracy_wisig}
        \begin{tikzpicture}
            \begin{axis}[
                width=0.95\columnwidth,
                height=0.52\columnwidth,
                xlabel={Number of rogue transmitters},
                ylabel={New-device accuracy (\%)},
                xmin=2,
                xmax=7,
                ymin=35,
                ymax=100,
                xtick={2,3,4,5,6,7},
                ytick={40,50,60,70,80,90,100},
                grid=major,
                tick label style={font=\footnotesize},
                label style={font=\footnotesize},
                legend style={
                    font=\scriptsize,
                    at={(0.38,0.223)},
                    anchor=north,
                    legend columns=2,
                    draw=black,
                    fill=white
                }
            ]

            \addplot+[mark=*, thick] coordinates {
                (2,90.0)
                (3,94.7)
                (4,84.0)
                (5,93.6)
                (6,90.0)
                (7,90.3)
            };
            \addlegendentry{With OSRWC}

            \addplot+[mark=square*, dashed, thick] coordinates {
                (2,90.0)
                (3,94.7)
                (4,79.0)
                (5,93.6)
                (6,40.0)
                (7,88.0)
            };
            \addlegendentry{Without OSRWC}

            \end{axis}
        \end{tikzpicture}
    }
    \vspace{0.3cm}
    \caption{
        \ac{osrwc} effect on new-device accuracy vs. $|\mySet{I}_{\rm R}|$.
    }
    \label{fig:new_device_accuracy_ablation_combined}
\end{figure}
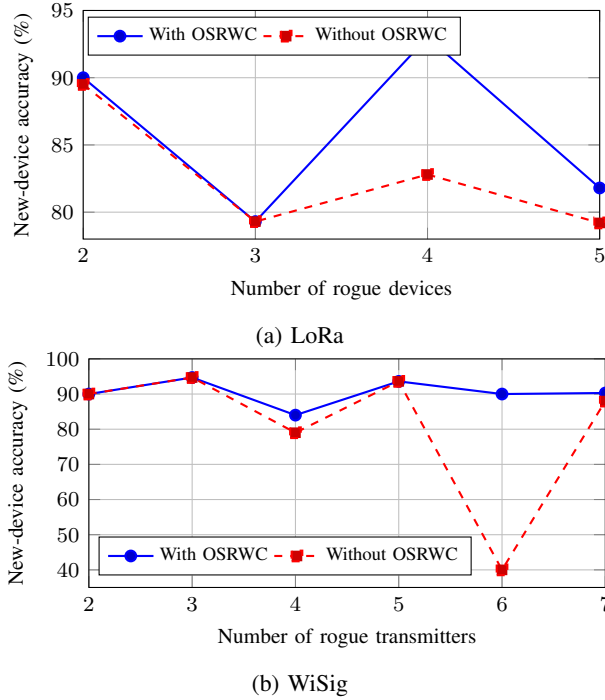

\subsection{Holistic Information-Transfer Ablation}
\label{subsec:osrwc_ablation}
Finally, we  evaluate the key holistic component of our method: transferring the  \ac{osr} score to  \ac{ncd}. We compare the  method with a configuration in which \ac{osrwc} is disabled. Thus, both variants use the same feature representation, the same registered-device model, the same binary open-set decision, and the same density-based discovery mechanism; they differ  in whether  the  \ac{osr} score is retained.
Fig.~\ref{fig:new_device_accuracy_ablation_combined} reports new-device accuracy on both datasets. On LoRa, the largest gain occurs with four rogue devices, where transferring the \ac{osr} information increases accuracy from $82.8\%$ to $93\%$; with five devices, the improvement is from $79.2\%$ to $81.8\%$. On WiSig, the effect becomes particularly pronounced in challenging multi-device settings. With six rogue transmitters, the unweighted discovery procedure attains only $40\%$ new-device accuracy, whereas \ac{osrwc} achieves $90\%$. With seven rogue transmitters, the corresponding improvement is from $88\%$ to $90.3\%$.

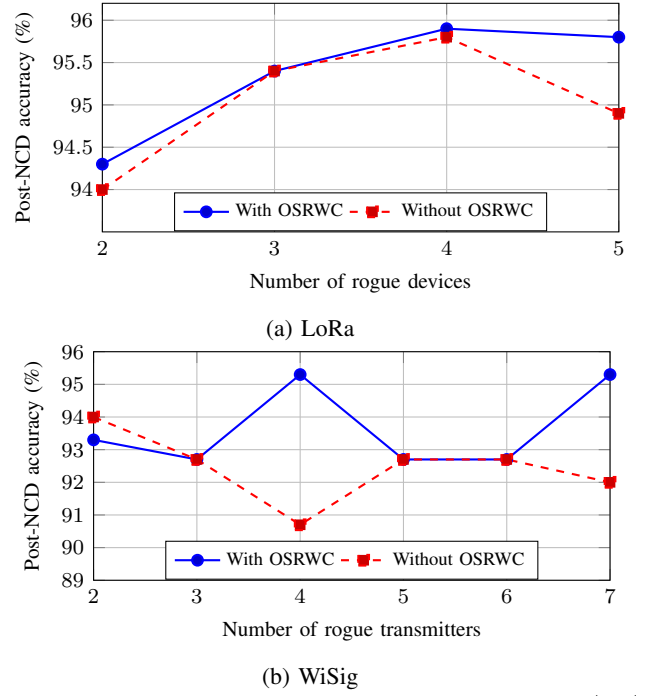
\begin{figure}
    \centering

    \subfloat[LoRa]{
        \label{fig:post_ncd_accuracy_lora}
        \begin{tikzpicture}
            \begin{axis}[
                width=0.95\columnwidth,
                height=0.52\columnwidth,
                xlabel={Number of rogue devices},
                ylabel={Post-\ac{ncd} accuracy (\%)},
                xmin=2,
                xmax=5,
                ymin=93.5,
                ymax=96.2,
                xtick={2,3,4,5},
                ytick={94.0,94.5,95.0,95.5,96.0},
                grid=major,
                tick label style={font=\footnotesize},
                label style={font=\footnotesize},
                legend style={
                    font=\scriptsize,
                    at={(0.5,0.173)},
                    anchor=north,
                    legend columns=2,
                    draw=black,
                    fill=white
                }
            ]

            \addplot+[mark=*, thick] coordinates {
                (2,94.3)
                (3,95.4)
                (4,95.9)
                (5,95.8)
            };
            \addlegendentry{With OSRWC}

            \addplot+[mark=square*, dashed, thick] coordinates {
                (2,94.0)
                (3,95.4)
                (4,95.8)
                (5,94.9)
            };
            \addlegendentry{Without OSRWC}

            \end{axis}
        \end{tikzpicture}
    }

    \vspace{0.25cm}

    \subfloat[WiSig]{
        \label{fig:post_ncd_accuracy_wisig}
        \begin{tikzpicture}
            \begin{axis}[
                width=0.95\columnwidth,
                height=0.52\columnwidth,
                xlabel={Number of rogue transmitters},
                ylabel={Post-\ac{ncd} accuracy (\%)},
                xmin=2,
                xmax=7,
                ymin=89,
                ymax=96,
                xtick={2,3,4,5,6,7},
                ytick={89,90,91,92,93,94,95,96},
                grid=major,
                tick label style={font=\footnotesize},
                label style={font=\footnotesize},
                legend style={
                    font=\scriptsize,
                    at={(0.5,0.173)},
                    anchor=north,
                    legend columns=2,
                    draw=black,
                    fill=white
                }
            ]

            \addplot+[mark=*, thick] coordinates {
                (2,93.3)
                (3,92.7)
                (4,95.3)
                (5,92.7)
                (6,92.7)
                (7,95.3)
            };
            \addlegendentry{With OSRWC}

            \addplot+[mark=square*, dashed, thick] coordinates {
                (2,94.0)
                (3,92.7)
                (4,90.7)
                (5,92.7)
                (6,92.7)
                (7,92.0)
            };
            \addlegendentry{Without OSRWC}

            \end{axis}
        \end{tikzpicture}
    }
    \vspace{0.3cm}
    \caption{
    \ac{osrwc} effect on post-discovery  identification vs. $|\mySet{I}_{\rm R}|$.
    }
    \label{fig:post_ncd_accuracy_ablation_combined}
\end{figure}

The effect extends beyond  discovery assignments. Fig.~\ref{fig:post_ncd_accuracy_ablation_combined} shows the resulting post-discovery closed-set accuracy. While the two configurations are comparable in the easier settings, \ac{osrwc} substantially improves the  system whenever unweighted clustering fails to provide sufficient support for one or more unseen-device regions. For example, on WiSig it increases post-discovery accuracy from $90.7\%$ to $95.3\%$ with four rogue transmitters and from $92.0\%$ to $95.3\%$ with seven. These results  demonstrate the benefit of the holistic formulation: information generated during \ac{osr} can improve the subsequent discovery task and the expanded identification system.

\begin{figure}
    \centering
    \begin{tikzpicture}
        \begin{axis}[
            width=\columnwidth,
            height=\figHeight,
            xlabel={SNR (dB)},
            ylabel={Rogue mapping accuracy (\%)},
            xmin=5,
            xmax=30,
            x dir=reverse,
            ymin=0,
            ymax=100,
            xtick={5,8,10,12,15,20,25,30},
            ytick={0,20,40,60,80,100},
            grid=major,
            tick label style={font=\footnotesize},
            label style={font=\footnotesize},
            legend style={
                font=\scriptsize,
                at={(0.97,0.97)},
                anchor=north east,
                draw=black,
                fill=white
            }
        ]

        \addplot+[
            mark=*,
            thick,
            error bars/.cd,
            y dir=both,
            y explicit
        ] coordinates {
            (30,88.2) +- (0,2.4)
            (25,88.9) +- (0,2.1)
            (20,79.0) +- (0,10.2)
            (15,78.1) +- (0,10.4)
            (12,66.2) +- (0,11.1)
            (10,50.5) +- (0,8.6)
            (8,17.5) +- (0,8.2)
            (5,7.1) +- (0,2.4)
        };
        \addlegendentry{With OSRWC}

        \addplot+[
            mark=square*,
            dashed,
            thick,
            error bars/.cd,
            y dir=both,
            y explicit
        ] coordinates {
            (30,88.8) +- (0,0.5)
            (25,87.9) +- (0,0.4)
            (20,62.1) +- (0,20.6)
            (15,36.0) +- (0,24.8)
            (12,30.3) +- (0,8.1)
            (10,25.3) +- (0,5.7)
            (8,0.0) +- (0,0.0)
            (5,0.0) +- (0,0.0)
        };
        \addlegendentry{Without OSRWC}

        \end{axis}
    \end{tikzpicture}
    \caption{
        \ac{osrwc} effect on rogue  accuracy vs. \ac{snr}, WiSig. 
    }
    \label{fig:osrwc_snr_mapping}
\end{figure}
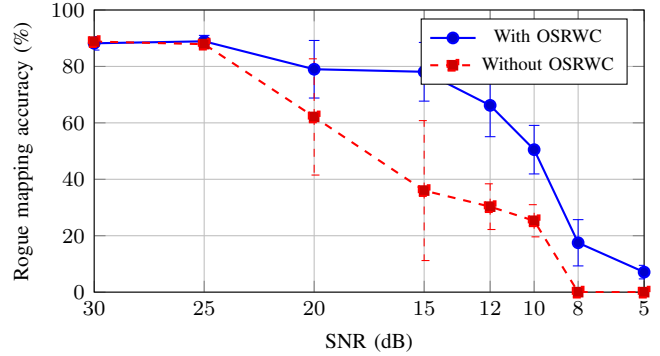

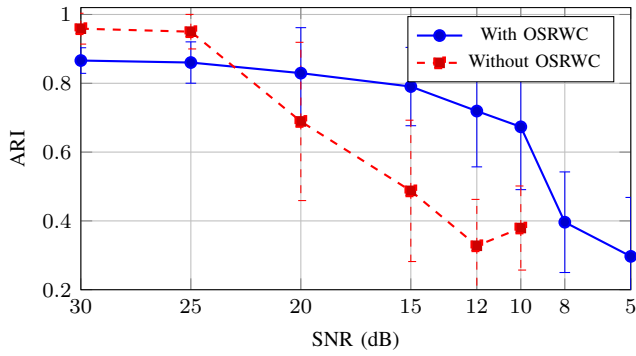
\begin{figure}
    \centering
    \begin{tikzpicture}
        \begin{axis}[
            width=\columnwidth,
            height=\figHeight,
            xlabel={SNR (dB)},
            ylabel={\ac{ari}},
            xmin=5,
            xmax=30,
            x dir=reverse,
            ymin=0.2,
            ymax=1.02,
            xtick={5,8,10,12,15,20,25,30},
            ytick={0.2,0.4,0.6,0.8,1.0},
            grid=major,
            tick label style={font=\footnotesize},
            label style={font=\footnotesize},
            legend style={
                font=\scriptsize,
                at={(0.97,0.97)},
                anchor=north east,
                draw=black,
                fill=white
            }
        ]

        \addplot+[
            mark=*,
            thick,
            error bars/.cd,
            y dir=both,
            y explicit
        ] coordinates {
            (30,0.8661) +- (0,0.0372)
            (25,0.8603) +- (0,0.0601)
            (20,0.8295) +- (0,0.1322)
            (15,0.7905) +- (0,0.1138)
            (12,0.7193) +- (0,0.1622)
            (10,0.6734) +- (0,0.1825)
            (8,0.3962) +- (0,0.1462)
            (5,0.2969) +- (0,0.1713)
        };
        \addlegendentry{With OSRWC}

        \addplot+[
            mark=square*,
            dashed,
            thick,
            error bars/.cd,
            y dir=both,
            y explicit
        ] coordinates {
            (30,0.9587) +- (0,0.0448)
            (25,0.9500) +- (0,0.0502)
            (20,0.6891) +- (0,0.2300)
            (15,0.4873) +- (0,0.2054)
            (12,0.3277) +- (0,0.1348)
            (10,0.3793) +- (0,0.1222)
        };
        \addlegendentry{Without OSRWC}

        \end{axis}
    \end{tikzpicture}
    \caption{
              \ac{osrwc} effect on \ac{ari} vs. \ac{snr}, WiSig.
        Without OSRWC   no clusters
        are discovered at $8$ and $5$ dB.
    }
    \label{fig:osrwc_snr_ari}
\end{figure}

We further examine the contribution of \ac{osrwc} under channel
degradation in Figs.~\ref{fig:osrwc_snr_mapping}--\ref{fig:osrwc_snr_ari}. At high \ac{snr}, both configurations achieve comparable rogue-device mapping accuracy,  as information transfer is not essential when the latent clusters are well separated. As the \ac{snr} decreases, the advantage of \ac{osrwc} becomes  pronounced. At $15$ dB, \ac{osrwc} achieves a mapping accuracy of $78.1\%$ and an \ac{ari} of $0.7905$, compared with only $36.0\%$ and $0.4873$, respectively, without information sharing. The gap persists at lower \ac{snr}: at $10$ dB, \ac{osrwc} still maps $50.5\%$ of the rogue samples with an \ac{ari} of $0.6734$, whereas the unweighted configuration reaches only $25.3\%$ mapping accuracy and an \ac{ari} of $0.3793$. At $8$ and $5$ dB, the unweighted discovery stage fails to form any clusters at all. These results indicate that transferring the continuous \ac{osr} score is particularly valuable when noise weakens the separation of unseen-device samples.


\section{Conclusions}
\label{sec:conclusion} 

We proposed a holistic open-world \ac{rffi} framework that unifies closed-set identification, \ac{osr}, and \ac{ncd} through an expandable vector-quantized latent representation. By transferring  open-set information to the discovery stage and incorporating newly identified devices through codebook expansion, our approach couples the different \ac{rffi} tasks without requiring retraining. Experiments on LoRa and WiSig demonstrate accurate identification and open-set detection with effective discovery and registration of unseen devices. 


\bibliographystyle{IEEEtran}
\bibliography{IEEEabrv,mybib}

\begin{thebibliography}{10}
\providecommand{\url}[1]{#1}
\csname url@samestyle\endcsname
\providecommand{\newblock}{\relax}
\providecommand{\bibinfo}[2]{#2}
\providecommand{\BIBentrySTDinterwordspacing}{\spaceskip=0pt\relax}
\providecommand{\BIBentryALTinterwordstretchfactor}{4}
\providecommand{\BIBentryALTinterwordspacing}{\spaceskip=\fontdimen2\font plus
\BIBentryALTinterwordstretchfactor\fontdimen3\font minus \fontdimen4\font\relax}
\providecommand{\BIBforeignlanguage}[2]{{%
\expandafter\ifx\csname l@#1\endcsname\relax
\typeout{** WARNING: IEEEtran.bst: No hyphenation pattern has been}%
\typeout{** loaded for the language `#1'. Using the pattern for}%
\typeout{** the default language instead.}%
\else
\language=\csname l@#1\endcsname
\fi
#2}}
\providecommand{\BIBdecl}{\relax}
\BIBdecl

\bibitem{hazan2026rf}
O.~Hazan, G.~Paryanti, and N.~Shlezinger, ``{RF} fingerprint identification with generalized class discovery via learning-aided vector quantization,'' in \emph{Proc. IEEE SPAWC}, 2026.

\bibitem{saadouni2025identification}
C.~Saadouni \emph{et~al.}, ``Identification techniques in the internet of things: Survey, taxonomy and research frontier,'' \emph{{IEEE} Commun. Surveys Tuts.}, vol.~28, pp. 593--632, 2025.

\bibitem{zhang2023radio}
J.~Zhang, G.~Shen, W.~Saad, and K.~Chowdhury, ``Radio frequency fingerprint identification for device authentication in the internet of things,'' \emph{{IEEE} Commun. Mag.}, vol.~61, no.~10, pp. 110--115, 2023.

\bibitem{zhang2021radio}
J.~Zhang \emph{et~al.}, ``Radio frequency fingerprint identification for narrowband systems, modelling and classification,'' \emph{{IEEE} Trans. Inf. Forensics Security}, vol.~16, pp. 3974--3987, 2021.

\bibitem{xu2016device}
Q.~Xu, R.~Zheng, W.~Saad, and Z.~Han, ``Device fingerprinting in wireless networks: Challenges and opportunities,'' \emph{{IEEE} Commun. Surveys Tuts.}, vol.~18, no.~1, pp. 94--104, 2015.

\bibitem{peng2018design}
L.~Peng, A.~Hu, J.~Zhang, Y.~Jiang, J.~Yu, and Y.~Yan, ``Design of a hybrid {RF} fingerprint extraction and device classification scheme,'' \emph{{IEEE} Internet Things J.}, vol.~6, no.~1, pp. 349--360, 2018.

\bibitem{jian2021radio}
T.~Jian \emph{et~al.}, ``Radio frequency fingerprinting on the edge,'' \emph{{IEEE} Trans. Mobile Comput.}, vol.~21, no.~11, pp. 4078--4093, 2021.

\bibitem{abbas2024radio}
S.~Abbas \emph{et~al.}, ``Radio frequency fingerprinting techniques for device identification: a survey,'' \emph{Int. J. Inf. Secur.}, vol.~23, no.~2, pp. 1389--1427, 2024.

\bibitem{he2025category}
Z.~He, Y.~Liu, and K.~Han, ``Category discovery: An open-world perspective,'' \emph{arXiv preprint arXiv:2509.22542}, 2025.

\bibitem{brik2008wireless}
V.~Brik, S.~Banerjee, M.~Gruteser, and S.~Oh, ``Wireless device identification with radiometric signatures,'' in \emph{ACM MobiCom}, 2008.

\bibitem{hall2003detection}
J.~Hall, M.~Barbeau, and E.~Kranakis, ``Detection of transient in radio frequency fingerprinting using signal phase,'' \emph{Wireless and Optical Communications}, vol.~9, p.~13, 2003.

\bibitem{huang2017radio}
Y.~Huang \emph{et~al.}, ``Radio frequency fingerprint extraction of radio emitter based on {I/Q} imbalance,'' \emph{Procedia Comput. Sci}, vol. 107, pp. 472--477, 2017.

\bibitem{candore2009robust}
A.~Candore, O.~Kocabas, and F.~Koushanfar, ``Robust stable radiometric fingerprinting for wireless devices,'' in \emph{IEEE HOST}, 2009.

\bibitem{patel2015non}
H.~Patel, ``Non-parametric feature generation for {RF}-fingerprinting on {ZigBee} devices,'' in \emph{IEEE CISDA}, 2015.

\bibitem{sankhe2019oracle}
K.~Sankhe \emph{et~al.}, ``{ORACLE}: Optimized radio classification through convolutional neural networks,'' in \emph{Proc. IEEE INFOCOM}, 2019.

\bibitem{shen2021lora}
G.~Shen, J.~Zhang, A.~Marshall, L.~Peng, and X.~Wang, ``Radio frequency fingerprint identification for {LoRa} using deep learning,'' \emph{{IEEE} J. Sel. Areas Commun.}, vol.~39, no.~8, pp. 2604--2616, 2021.

\bibitem{hu2025lightweight}
Y.~Hu, Y.~Fu, and Y.~Chen, ``A lightweight {LSTM}-based open-set {RF} fingerprinting identification for edge deployment,'' \emph{Scientific Reports}, vol.~15, no.~1, p. 41568, 2025.

\bibitem{wang2022few}
Y.~Wang \emph{et~al.}, ``Few-shot specific emitter identification via deep metric ensemble learning,'' \emph{{IEEE} Internet Things J.}, vol.~9, no.~24, pp. 24\,980--24\,994, 2022.

\bibitem{wu2023specific}
Z.~Wu, F.~Wang, and B.~He, ``Specific emitter identification via contrastive learning,'' \emph{{IEEE} Commun. Lett.}, vol.~27, no.~4, pp. 1160--1164, 2023.

\bibitem{yao2023few}
Z.~Yao \emph{et~al.}, ``Few-shot specific emitter identification using asymmetric masked auto-encoder,'' \emph{{IEEE} Commun. Lett.}, vol.~27, no.~10, pp. 2657--2661, 2023.

\bibitem{liu2023overcoming}
C.~Liu \emph{et~al.}, ``Overcoming data limitations: A few-shot specific emitter identification method using self-supervised learning and adversarial augmentation,'' \emph{{IEEE} Trans. Inf. Forensics Security}, vol.~19, pp. 500--513, 2023.

\bibitem{guo2024toward}
L.~Guo, C.~Liu, Y.~Liu, Y.~Lin, and G.~Gui, ``Toward open-set specific emitter identification using auxiliary classifier generative adversarial network and openmax,'' \emph{{IEEE} Trans. on Cogn. Commun. Netw.}, vol.~10, no.~6, pp. 2019--2028, 2024.

\bibitem{zhang2025open}
T.~Zhang \emph{et~al.}, ``An open-set few-shot class incremental learning framework for specific emitter identification,'' \emph{{IEEE} Trans. on Cogn. Commun. Netw.}, vol.~12, pp. 602--616, 2025.

\bibitem{li2025fscil}
D.~Li, X.~Chen, S.~Hong, J.~Qi, and H.~Sun, ``{FSCIL-SEI}: Few-shot class-incremental learning approach for specific emitter identification,'' \emph{{IEEE} Trans. Instrum. Meas.}, vol.~74, 2025.

\bibitem{xie2025novel}
W.~Xie, H.~Wang, Z.~Shen, X.~Li, Z.~Liu, and H.~Jiang, ``A novel radio frequency fingerprint identification scheme for few-shot open-set recognition,'' \emph{{IEEE} Internet Things J.}, vol.~12, no.~13, pp. 25\,691--25\,706, 2025.

\bibitem{cao2021open}
A.~Cao, Y.~Luo, and D.~Klabjan, ``Open-set recognition with {G}aussian mixture variational autoencoders,'' in \emph{Proceedings of the AAAI Conference on Artificial Intelligence}, vol.~35, no.~8, 2021, pp. 6877--6884.

\bibitem{ma2025open}
Z.~Ma, S.~Fang, and Y.~Fan, ``Open-set radio frequency fingerprint identification method based on multi-task prototype learning,'' \emph{Sensors}, vol.~25, no.~17, p. 5415, 2025.

\bibitem{cai2025open}
D.~Cai \emph{et~al.}, ``Open set {RF} fingerprinting identification: A joint prediction and siamese comparison framework,'' in \emph{Proc. IEEE ICC}, 2025.

\bibitem{luo2025radio}
M.~Luo, X.~Liu, X.~Di, and H.~Qi, ``Radio frequency fingerprint recognition method based on feature enhancement and selective channel-spatial kernel,'' \emph{Computer Networks}, p. 111574, 2025.

\bibitem{karunaratne2021open}
S.~Karunaratne, S.~Hanna, and D.~Cabric, ``Open set {RF} fingerprinting using generative outlier augmentation,'' in \emph{IEEE GLOBECOM}, 2021.

\bibitem{vaze2022gcd}
S.~Vaze, K.~Han, A.~Vedaldi, and A.~Zisserman, ``Generalized category discovery,'' in \emph{IEEE CVPR}, 2022.

\bibitem{wen2023parametric}
X.~Wen, B.~Zhao, and X.~Qi, ``Parametric classification for generalized category discovery: A baseline study,'' in \emph{IEEE ICCV}, 2023.

\bibitem{fishel2025remote}
E.~Fishel, M.~Malka, S.~Ginzach, and N.~Shlezinger, ``Remote inference over dynamic links via adaptive rate deep task-oriented vector quantization,'' \emph{{IEEE} Trans. Signal Process.}, vol.~73, pp. 3557--3571, 2025.

\bibitem{chen2024information}
J.~Chen \emph{et~al.}, ``Information compression in the {AI} era: Recent advances and future challenges,'' \emph{{IEEE} J. Sel. Areas Commun.}, vol.~43, no.~7, pp. 2333--2348, 2025.

\bibitem{hanna2022wisig}
S.~Hanna, S.~Karunaratne, and D.~Cabric, ``Wi{S}ig: A large-scale {WiFi} signal dataset for receiver and channel agnostic {RF} fingerprinting,'' \emph{{IEEE} Access}, vol.~10, pp. 22\,808--22\,818, 2022.

\bibitem{polyanskiy2014lecture}
Y.~Polyanskiy and Y.~Wu, ``Lecture notes on information theory,'' \emph{Lecture Notes for 6.441 (MIT) and STAT 664 (Yale)}, 2012-2017.

\bibitem{linde1980algorithm}
Y.~Linde, A.~Buzo, and R.~Gray, ``An algorithm for vector quantizer design,'' \emph{{IEEE} Trans. Commun.}, vol.~28, no.~1, pp. 84--95, 1980.

\bibitem{van2017neural}
A.~Van Den~Oord and O.~Vinyals, ``Neural discrete representation learning,'' \emph{Advances in Neural Information Processing Systems}, vol.~30, 2017.

\bibitem{fifty2024restructuring}
C.~Fifty \emph{et~al.}, ``Restructuring vector quantization with the rotation trick,'' in \emph{International Conference on Learning Representations (ICLR)}, 2025.

\bibitem{malka2024decentralized}
M.~Malka, E.~Farhan, H.~Morgenstern, and N.~Shlezinger, ``Decentralized low-latency collaborative inference via ensembles on the edge,'' \emph{{IEEE} Trans. Wireless Commun.}, vol.~24, no.~1, pp. 598--614, 2025.

\bibitem{huang2023towards}
M.~Huang \emph{et~al.}, ``Towards accurate image coding: Improved autoregressive image generation with dynamic vector quantization,'' in \emph{IEEE CVPR}, 2023.

\bibitem{ester1996density}
M.~Ester \emph{et~al.}, ``A density-based algorithm for discovering clusters in large spatial databases with noise,'' \emph{KDD}, 1996.

\bibitem{campello2013density}
R.~J.~G. Campello, D.~Moulavi, and J.~Sander, ``Density-based clustering based on hierarchical density estimates,'' in \emph{Advances in Knowledge Discovery and Data Mining}.\hskip 1em plus 0.5em minus 0.4em\relax Springer, 2013, pp. 160--172.

\bibitem{khosla2020supervised}
P.~Khosla, P.~Teterwak, C.~Wang, A.~Sarna, Y.~Tian, P.~Isola, A.~Maschinot, C.~Liu, and D.~Krishnan, ``Supervised contrastive learning,'' in \emph{Advances in Neural Information Processing Systems}, vol.~33, 2020, pp. 18\,661--18\,673.

\bibitem{shen2022towards}
G.~Shen, J.~Zhang, A.~Marshall, and J.~R. Cavallaro, ``Towards scalable and channel-robust radio frequency fingerprint identification for {LoRa},'' \emph{{IEEE} Trans. Inf. Forensics Security}, vol.~17, pp. 774--787, 2022.

\bibitem{zhao2024anomaly}
T.~Zhao, S.~Sarkar, Y.~Tian, and D.~Cabric, ``Anomaly transmitter recognition and tracking,'' in \emph{IEEE International Symposium on Dynamic Spectrum Access Networks (DySPAN)}, 2024, pp. 357--364.

\bibitem{gao2023opengcdassistingopenworld}
F.~Gao, W.~Zhong, Z.~Cao, X.~Peng, and Z.~Li, ``Open{GCD}: Assisting open world recognition with generalized category discovery,'' \emph{arXiv preprint arXiv:2308.06926}, 2023.

\bibitem{liu2023openworldsemisupervisednovelclass}
J.~Liu, Y.~Wang, T.~Zhang, Y.~Fan, Q.~Yang, and J.~Shao, ``Open-world semi-supervised novel class discovery,'' \emph{arXiv preprint arXiv:2305.13095}, 2023.

\end{thebibliography}

\end{document}